\documentclass[aps,prc,superscriptaddress,twoside,twocolumn,nofootinbib,showpacs]{revtex4}
\usepackage{amsmath,amssymb}
\usepackage{graphicx}
\newcommand{\fs}[1]{{#1\!\!\!/}}
\newcommand{\hc}{\text{h.\,c.}}
\begin{document}

\title{Consistent analysis of the reactions
$\boldsymbol{\gamma p\to p \eta^\prime}$ and $\boldsymbol{pp\to pp\eta'}$}
\author{K. Nakayama}
\affiliation{Department of Physics and Astronomy, University of Georgia, Athens, GA 30602, USA}
\affiliation{Institut f{\"u}r Kernphysik (Theorie),
Forschungszentrum J{\"u}lich,
D-52425 J{\"u}lich, Germany}
\author{H. Haberzettl}
\affiliation{Center for Nuclear Studies,
Department of Physics,
The George Washington University,
Washington, DC 20052, USA}
\affiliation{Institut f{\"u}r Kernphysik (Theorie),
Forschungszentrum J{\"u}lich,
D-52425 J{\"u}lich, Germany}

\date{13 January 2004}

\begin{abstract}
The production of $\eta'$ mesons in the reactions $\gamma p\to p\eta'$ and $pp\to
pp\eta'$ is described consistently within a relativistic meson exchange model of hadronic
interactions. The photoproduction can be described quite well over the entire energy
range of available data by considering an $S_{11}$ and a $P_{11}$ resonance, in addition
to the $t$-channel mesonic current. The observed angular distribution is due to the
interference between the $t$-channel and the nucleon resonance $s$- and $u$-channel
contributions. Our analysis yields positions close to 1650 MeV and 1870 MeV for the
$S_{11}$ and $P_{11}$ resonances, respectively. We argue that, at present, identifying
these states with the known $S_{11}(1650)$ resonance and the missing $P_{11}$ resonance
predicted at 1880 MeV, respectively, would be premature. It is found that the nucleonic
current is relatively small and that the $NN\eta^\prime$ coupling constant cannot be much
larger than $g_{NN\eta^\prime}=3$. As for the $p p \to p p \eta^\prime$ reaction,
different current contributions are constrained by a combined analysis of this and the
photoproduction reaction. Difficulties to simultaneously account for the 47-MeV and
144-MeV angular distributions measured by the COSY-11 and DISTO collaborations,
respectively, are addressed.
\end{abstract}


\pacs{25.20.Lj, 
      13.60.Le, 
      13.75.-n,  
      14.20.Gk 
      \hfill {\bfseries\footnotesize[nucl-th/0401030---PRC (05/2004)]}%
      } %

\maketitle

\section{Introduction}

The study of the intrinsic properties of the $\eta^\prime$ meson as well as its
production processes in elementary particle and hadron physics is of particular
interest for various reasons. The properties of $\eta^\prime$ are largely
governed by the dynamics of the QCD $U_A(1)$ axial vector anomaly
\cite{UA1,Weinberg,'tHooft1,Witten,Venez,Christ}.
Expressed in the (pseudoscalar) quark-flavor basis, the physically observed
$\eta$ and $\eta^\prime$ mesons may be written as
\begin{equation}
\begin{pmatrix} \eta \\ \eta'\end{pmatrix}
  =\begin{pmatrix} \cos\alpha & -\sin\alpha \\ \sin\alpha & \cos\alpha \end{pmatrix}
\begin{pmatrix} \eta_q \\ \eta_s\end{pmatrix}~,
\label{etaetapmixqs}
\end{equation}
where $\eta_s \equiv {s\bar s}$ and $\eta_q \equiv (u\bar u + d\bar d)/\sqrt{2}$ describe
the strange and nonstrange quark-antiquark states, respectively. The $U_A(1)$ anomaly
mediates $\eta_q$--$\eta_s$ transitions and therefore plays a central role in
understanding the  $\eta$--$\eta^\prime$ mixing \cite{Feldmann1}. The mixing angle
$\alpha$ is shown to be fairly constant \cite{Kaiser,Feldmann1}, and a weighted average
value of $\alpha = 39.3^\circ \pm 1.0^\circ$ has been extracted \cite{Feldmann2}. Quite
recently, the KLOE collaboration \cite{KLOE} has reported a value of $\alpha =
41.8^\circ{^{+1.9^\circ}_{-1.6^\circ}}$. As can be seen from Eq.~(\ref{etaetapmixqs}),
such a value of the mixing angle results in a considerable amount of $s\bar s$ in both
the $\eta$ and $\eta^\prime$ mesons. By contrast, the corresponding mixing angle  for the
vector mesons $\omega$ and $\phi$ is quite small ($\approx 3.4^\circ$
\cite{omegaphimixang1,PDG,omegaphimixang2}), providing an $\omega$ with nearly no $s\bar
s$ and a $\phi$ being almost a pure $s\bar s$ state.

Therefore, instead of using the vector mesons $\omega$ and $\phi$, production processes
involving $\eta^\prime$ and $\eta$ offer an alternative way of probing the strangeness
content of the nucleon. Due to the fact that both $\eta$ and $\eta^\prime$ contain a
significant amount of $s\bar s$, but of opposite phase with respect to the nonstrange
$u\bar u + d\bar d$ component, significant interference effects involving the
strange-quark piece of the nucleon wave function are possible \cite{Dover}. In fact, it
has been proposed that the relative cross sections for the reactions induced by
pseudoscalar mesons, $\eta p, \eta^\prime p \rightarrow \eta p, \eta^\prime p,
K^+\Lambda$ and $\pi^- p \rightarrow \eta n, \eta^\prime n$, provide a sensitive test for
the presence of the $s\bar s$ component in the nucleon wave function \cite{Dover}.

Due to its nontrivial properties, the QCD vacuum exhibits strong gluonic fluctuations
with pseudoscalar quantum numbers to which the $\eta_{q,s}$ states can couple via the
$U_A(1)$ axial anomaly. The nonperturbative gluon dynamics and the axial anomaly
\cite{'tHooft1,Witten,Venez,'tHooft2} are thought to be responsible for the generation of
the much larger mass of $\eta^\prime$ as compared to the masses of other members of the
$SU(3)$ pseudoscalar meson nonet known as the Goldstone bosons. The masses of the Goldstone
bosons are generated by the spontaneous breaking of chiral symmetry
\cite{Weinberg,Christ,GBmass}. The $\eta^\prime$ meson is, therefore, thought to couple
strongly to gluons via the $U_A(1)$ axial anomaly coupling \cite{Ball,Gluons}. The
unexpectedly large branching ratio measured recently for the inclusive decay of beauty
particles, $B \rightarrow \eta ^{\prime }+X$ \cite{CLEO}, has been interpreted as possible
experimental evidence in this respect \cite{Gluonic}. To date, the KLOE collaboration has
recently found that the gluonium content in the $\eta^\prime$ is consistent with a
fraction below $15\%$ \cite{KLOE}. In any case, if there is a strong coupling of
$\eta^\prime$ meson to gluons, it would be conceivable that short-range reaction
processes such as $pp \rightarrow pp\eta^\prime$ might reveal the gluonic degrees of
freedom in the low energy interactions involving nucleons and $\eta^\prime$
\cite{ppetapgluon}.

One of the properties of the $\eta ^{\prime }$ meson of extreme importance is its yet
poorly known coupling strength to the nucleon. This has attracted much attention in
connection with the so-called ``nucleon-spin crisis'' in polarized deep inelastic lepton
scattering \cite{EMC88}. In the zero-momentum limit, the $NN\eta ^{\prime }$ coupling
constant $g_{NN\eta ^{\prime }}$ is related to the flavor singlet axial charge $G_A$
through the flavor singlet Goldberger-Treiman relation \cite{Shore} (see also
Refs.\cite{Efremov,Venez1})
\begin{equation}
2m_N\,G_A(0) = F\,g_{NN\eta^\prime}(0) +
{\frac {F^2}{2N_F}}\,m_{\eta^\prime}\,g_{NNG}(0)~,
\label{spinfrac}
\end{equation}
where $F\sim \sqrt{2N_F}F_\pi$ is a renormalization-group invariant decay constant; $N_F$
and $F_\pi$ denote the number of flavors and the pion decay constant, respectively.
$g_{NNG}$ describes the coupling of the nucleon to the gluons arising from contributions
violating the Okubo-Zweig-Iizuka rule \cite{OZI}. The EMC collaboration \cite{EMC88} has
measured an unexpectedly small value of $G_A(0) \sim 0.20$--$0.35$. The first term on the
right-hand side of the above equation corresponds to the quark contribution to the ``spin''
of the proton, and the second term to the gluon contribution \cite{Venez1,x1}. Therefore,
if $g_{NN\eta'}(0)$ is known, Eq.~(\ref{spinfrac}) may be used to extract the coupling
$g_{NNG}(0)$. However, unfortunately, there is no direct experimental measurement of
$g_{NN\eta'}(0)$ so far. Reaction processes where the $\eta^\prime$ meson is produced
directly off a nucleon, such as $\gamma p \rightarrow p\eta^\prime$ and $pp \rightarrow
pp\eta^\prime$, may thus offer a unique opportunity to extract this coupling constant.
Of course, other production mechanisms, such as meson exchange and nucleon resonance
currents, must be taken into account before a quantitative determination of
$g_{NN\eta^\prime}$ is possible.

Yet another interesting aspect in studying $\eta^\prime$ production processes is that
they may provide an alternative tool to extract information on nucleon resonances, $N^*$.
Current knowledge of most of the nucleon resonances is mainly due to the study of $\pi N$
scattering and/or pion photoproduction off the nucleon. Reaction processes such as
$\eta^\prime$ photoproduction provide opportunities to study those resonances that couple
only weakly to pions, especially, in the less explored higher $N^*$ mass region of
``missing resonances'' \cite{Capstick1}. Missing resonances are those predicted by
quark models but not found in more traditional pion-production reactions
\cite{Capstick1}.

In the present work, we concentrate on the reactions $\gamma p \rightarrow p\eta^\prime$
and $pp \rightarrow pp\eta^\prime$. So far there exists only a limited number of studies
of the $\eta^\prime$ photoproduction both experimentally \cite{ABBHHM,SAPHIR} and
theoretically \cite{Zhang,Li,Borasoy,Zhao}. Zhang \emph{et al.}~\cite{Zhang}, in their
theoretical investigation using an effective Lagrangian approach, have emphasized the
role of the $D_{13}(2080)$ resonance in the description of the, then, existing data
\cite{ABBHHM}, while Li~\cite{Li} has described those data within a constituent quark
model with the off-shell $S_{11}(1535)$ excitation as the dominant contribution. The
authors of Ref.~\cite{SAPHIR} described their data --- obtained with much higher
statistics than the previous measurements \cite{ABBHHM} --- in the energy region from
threshold to 2.6\,GeV under the assumption of resonance dominance. They considered an
$S_{11}$ and a $P_{11}$ resonance with extracted masses of 1897 and 1986\,MeV,
respectively. The former resonance was needed to explain the energy dependence of the
total cross section which exhibits a steep rise and falloff close to threshold. The
$P_{11}(1986)$ resonance was needed to account for the measured forward rising angular
distributions. In a calculation similar to that in Ref.~\cite{Li}, Zhao~\cite{Zhao}
introduced also a $P_{13}$ and an $F_{13}$ resonance to describe the SAPHIR data
\cite{SAPHIR}. In both these quark model calculations, no ($t$-channel) vector meson
exchange contribution was considered. Based on a $U(3)$ baryon chiral perturbation theory,
Borasoy~\cite{Borasoy} introduced the off-shell $P_{11}(1440)$ and $S_{11}(1535)$
resonances, in addition to the Born and vector meson exchange contributions, to describe
the data \cite{SAPHIR}. Quite recently, Chiang \emph{et al.}~\cite{Chiang} have put
forward a model for $\eta'$ photoproduction that considers the $t$-channel vector meson
exchanges in terms of Regge trajectories to comply with high energy behavior. In their
calculation, which was applied to the SAPHIR data~ \cite{SAPHIR} (that cover an energy
region $< 2.6$\,GeV), the interference of the Regge trajectories with an $S_{11}$
resonance is the underlying mechanism responsible for reproducing the data and no need of
any $P_{11}$ resonance contribution was found. In contrast, also in a quite recent
calculation, Sibirtsev \emph{et al.}~\cite{Sibirtsev} have described the SAPHIR data by
considering the $t$-channel $\rho$- and $\omega$-meson exchange contributions with an
exponential form factor at the $\gamma\eta^\prime v$ vertex ($v=\rho,\omega$). The
observed forward rise of the angular distribution is then largely accounted for by the
($t$-dependent) exponential form factor. In addition, the $S_{11}(1535)$ resonance is
introduced in order to account for the energy dependence of the total cross section.
Sibirtsev \emph{et al.}~\cite{Sibirtsev} have also speculated that the $\eta'$
photoproduction at high energies and large $t$ may be useful in determining the $NN\eta'$
coupling constant $g_{NN\eta'}$. New experimental investigations of $\eta^\prime$
photoproduction are currently being carried out at JLab by the CLAS collaboration
\cite{CLAS} and at ELSA by the Crystal Barrel collaboration \cite{CristalBarrel}.

The $pp\rightarrow pp\eta^\prime$ reaction has been a subject of increasing attention in
the last few years. Experimental data on total cross section exist for excess energies up
to $Q \sim 24$~\,MeV~\cite{ppetapdata}, in addition to the total cross section and the
angular distribution at $Q=143.8$\,MeV from the DISTO collaboration \cite{Balestra}. The
new total cross section data in the excess energy range of $Q=26$--47\,MeV and an
angular distribution at $Q=46.6$\,MeV have been just reported by the COSY-11
collaboration \cite{Khoukaz}, filling in partly the gap between the near threshold
\cite{ppetapdata} and higher energy DISTO data \cite{Balestra}. Theoretically, the
$pp\rightarrow pp\eta^\prime$ reaction has been investigated by a number of authors
\cite{ppetaptheory,Nak1} within meson-exchange approaches of varying degrees of
sophistication. In particular, in Ref.~\cite{Nak1}, we have explored the possible role of
the nucleonic, mesonic, and resonance current contributions. The $S_{11}(1987)$ and
$P_{11}(1986)$ resonances as determined by the SAPHIR collaboration \cite{SAPHIR} have
been considered for the resonance current. Due to the scarcity of the then available data
(total cross sections up to $Q \approx 10$\,MeV), it was not possible to quantitatively
constrain each of these currents. With the increase of the data base since then, we are
now in a much better position to learn about this reaction than was possible before.

\begin{figure}[t]
\includegraphics[width=\columnwidth,angle=0,clip]{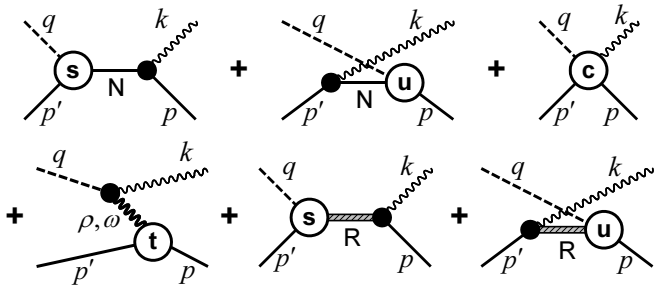}
\caption{\label{fig:etaphotoprod}%
Diagrams contributing to $\gamma p \to \eta' p$. Time proceeds from right to left. The
intermediate baryon states are denoted \textsf{N} for the nucleon, and \textsf{R} for
the $S_{11}$ and $P_{11}$ resonances. The intermediate mesons in the $t$-channel are
$\rho$ and $\omega$. The external legs are labeled by the four-momenta of the respective
particles and the labels \textsf{s}, \textsf{u}, and \textsf{t} of the hadronic vertices
correspond to the off-shell Mandelstam variables of the respective intermediate
particles. The three diagrams in the lower part of the diagram are transverse
individually; the three diagrams in the upper part are made gauge-invariant by an
appropriate choice (see text) of the contact current depicted in the top-right diagram.
The nucleonic current (nuc) referred to in the text corresponds to the top line of
diagrams; the meson-exchange current (mec) and resonance current contributions
correspond, respectively, to the leftmost diagram and the two diagrams on the right of
the bottom line of diagrams.}
\end{figure}

\begin{figure}[t]
\includegraphics[width=\columnwidth,angle=0,clip]{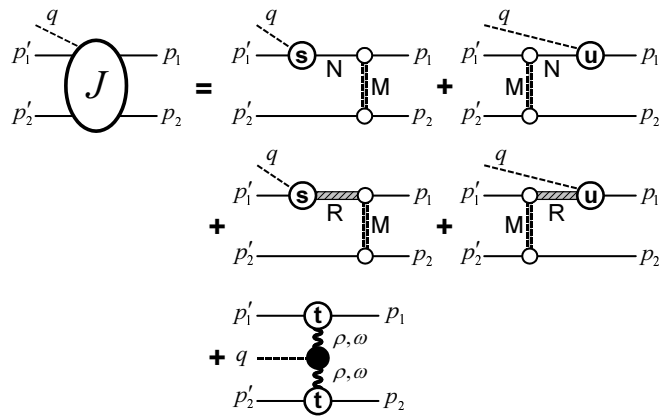}
\caption{\label{fig:ppetaprod}%
Basic production mechanisms for $pp\to pp \eta'$. Time proceeds from right to left. The
full amplitude, with additional initial- and final-state contributions, is given by
Eq.~(\ref{eq:ppetaprod}). As in Fig.~\ref{fig:etaphotoprod}, \textsf{N} and \textsf{R}
denote the intermediate nucleon and resonances, respectively, and \textsf{M} incorporates
all exchanges of mesons $\pi$, $\eta$, $\rho$, $\omega$, $\sigma$, and $a_0$
($\equiv\delta$) for the nucleon graphs and $\pi$, $\rho$, and $\omega$ for the resonance
graphs. External legs are labeled by the four-momenta of the respective particles; the
hadronic vertices \textsf{s}, \textsf{u}, and \textsf{t} here correspond to the same
kinematic situations, respectively, as those identified similarly in
Fig.~\ref{fig:etaphotoprod}. The nucleonic, resonance, and meson-exchange contributions
referred to in the text correspond, respectively, to the first, second, and third lines of
the diagrams on the right-hand side.}
\end{figure}

The major purpose of the present work is to perform a combined analysis of the $\gamma p
\rightarrow p\eta^\prime$ and $pp \rightarrow pp\eta^\prime$ reactions within a
relativistic meson-exchange model of hadronic interactions (see
Figs.~\ref{fig:etaphotoprod} and \ref{fig:ppetaprod}). For the $\eta^\prime$
photoproduction, in the $s$ and $u$ channels, we consider contributions due to the
intermediate nucleon and the nucleon resonances and  in the $t$-channel, we take into
account $\rho$ and $\omega$ meson exchanges. Since we employ the physical coupling
constants and physical masses for all intermediate particles in all the channels, the
$s$-channel diagrams also account for the pole part of the $N\eta^\prime$ final-state
interaction (FSI)~\cite{Pearce}. For the nonpole part of the FSI, the $u$ and
$t$ channels correspond to the Born approximation of the corresponding $N\eta'$
$T$-matrix. Phenomenological form factors are attached to each vertex in all channels.
The total amplitude is constrained to obey gauge invariance following the prescription of
Refs.~\cite{hh97g,hhtree98,dw2}. The photoproduction amplitude thus obtained is then used
in the construction of the basic $\eta'$ production amplitude in $pp \rightarrow
pp\eta^\prime$ by replacing the photons with relevant mesons which, in turn, are attached
to the second nucleon (see Fig.~\ref{fig:ppetaprod}). Hereafter, the basic $\eta^\prime$
production amplitude is referred to as the $\eta^\prime$ production current following the
nomenclature employed in Ref.~\cite{Nak1}. The $pp \rightarrow pp\eta^\prime$ reaction is
then described in a Distorted-Wave Born Approximation (DWBA) which includes both the
nucleon-nucleon ($NN$) final-state interaction and the initial-state interaction
(ISI).

The present paper is organized as follows. In Sec.~II our model for the $\gamma p
\rightarrow p\eta^\prime$ and $pp \rightarrow pp\eta^\prime$ reactions is described
briefly. The numerical results are discussed in Sec.~III, and in Sec.~IV we present our
summarizing conclusions.

\section{Formalism}

The dynamical content of our approach is summarized by the graphs of
Figs.~\ref{fig:etaphotoprod} and \ref{fig:ppetaprod}. For the $\eta'$ photoproduction, we
employ the tree graphs of Fig.~\ref{fig:etaphotoprod} with form factors at the vertices
to account for the hadronic structure. The gauge invariance of this production current is
ensured by adding a phenomenological contact current, according to the prescription of
Refs.~\cite{hh97g,hhtree98}. This contact term (see below) may also be thought of as
mocking up the neglected parts of the final-state interaction. The hadronic $pp\to pp\eta'$
reaction is described according to the model put forward in Refs.~\cite{Nak1,Nak2}. The
DWBA amplitude $M$ for this process is given by
\cite{Nak2}
\begin{equation}
M = (1 + T_f G_f) J (1 +  G_i T_i) \ ,
  \label{eq:ppetaprod}
\end{equation}
where $T_{n}$, with $n=i,f$, denotes the $NN$ $T$-matrix interaction in the initial ($i$)
or final ($f$) state, and $G_n$ is the corresponding two-nucleon propagator (which
absorbs the factor $i$ found in the DWBA formula given in Ref.~\cite{Nak2}).  $J$ sums up
the basic $\eta'$ production mechanisms depicted in Fig.~\ref{fig:ppetaprod}. In the
absence of models capable of providing a reliable off-shell $NN$ ISI, we consider it only
in the on-shell approximation following Ref.~\cite{Nak2}. This is shown to be a reasonable
approximation for calculating cross sections \cite{Hanhart}. The $NN$ FSI is generated by
using the Bonn potential \cite{Bonn}. We use the Blankenbecler-Sugar propagator for the
two-nucleon propagator $G_f$ in Eq.~(\ref{eq:ppetaprod}) in order to be consistent with
the $NN$ interaction used. The Coulomb force is ignored in the present calculation; its
effect is known to be relevant only in the energy region very close to threshold (excess
energies less than 5 MeV) \cite{Nak1}. In the present work we concentrate our attention
on the higher excess energy region where the Coulomb effect is negligible.

The interaction Lagrangian used to construct our model for the basic production
amplitudes is given below. For further convenience, we define the operators
\begin{equation}
\Gamma^{(+)} = \gamma_5
\qquad\text{and}\qquad
\Gamma^{(-)} = 1~.
\end{equation}

\subsection{Electromagnetic interaction Lagrangians}

The electromagnetic vertices are derived from the following Lagrangian densities.

\textit{$NN\gamma$ Lagrangian:}
\begin{equation}
{\cal L}_{NN\gamma} =  -e \bar N \left\{
\left( \gamma^\mu - \kappa_p\frac{\sigma^{\mu\nu}\partial_\nu}{2m_N} \right) A_\mu \right\} N ~,
\label{eq:gammaNN}
\end{equation}
where $A_\mu$ and $N$ stand for the photon and nucleon fields, respectively.
$m_N$ is the nucleon mass, $e$ the elementary charge unit,
and $\kappa_p=1.793$ the anomalous magnetic moment of the proton.

\textit{$NN^*\gamma$ Lagrangian (resonance $N^*=S_{11},P_{11}$):}
\begin{equation}
{\cal L}^{(\pm)}_{NN^*\gamma}  =  \frac {g_{NN^*\gamma}\,e} {m_{N^*}+m_N}\,
\bar N^* \Gamma^{(\mp)} \sigma_{\mu\nu}
(\partial^\nu A^\mu) N +\hc~,
\label{eq:gammaNNstar}
\end{equation}
where $N^*$ stands for the resonance field. The upper and lower signs on the left refer
to the even ($+$) and odd ($-$) parity resonances, respectively; $m_{N^*}$ is the
resonance mass and $g_{NN^{*}\gamma}$ the coupling constant. Both parameters are fit
parameters.

\textit{$\eta' v \gamma$ Lagrangian (vector meson $v=\rho,\omega$):}\\
\begin{equation}
{\cal L}_{\eta^\prime v \gamma} =  - \frac{g_{\eta^\prime v \gamma}e}{m_v}
\varepsilon_{\alpha\beta\nu\mu} (\partial^\alpha V^\beta)
(\partial^\nu A^\mu) \eta^\prime ~,
\label{eq:gammaVeta}
\end{equation}
where $\varepsilon^{\mu\nu\alpha\beta}$ is the Levi-Civita tensor. $V^\beta$ stands for
the vector meson field ($= \rho_3^\beta,\, \omega^\beta$). The resulting $\eta^\prime v
\gamma$ vertex is multiplied by the form factor $F_v(t)$ which describes the off-shell
behavior of the intermediate vector meson with squared momentum transfer $t=(p-p')^2$
(cf.\ fourth diagram in Fig.~\ref{fig:etaphotoprod}). In general, we use the dipole form
\begin{equation}
  F_v(t) = \left( \frac{\Lambda_v^2-m_v^2}{\Lambda_v^2-t} \right)^2
  \label{eq:dipoleFF}
\end{equation}
(see, however, Fig.~\ref{fig:e_mr2} below and its discussion in the text). The cutoff
$\Lambda_v$, taken to be identical for both $\rho$ and $\omega$, is a fit parameter. The
coupling constants $g_{\eta' v \gamma}$ in Eq.(\ref{eq:gammaVeta}) are taken from
radiative decays \cite{PDG}; their signs are inferred from $SU(3)$ symmetry
considerations following Ref.~\cite{Nak1} in conjunction with the sign of the coupling
constant $g_{\pi v \gamma}$ determined from a study of pion photoproduction in the 1
GeV energy region \cite{Garcilazo}.

\subsection{Hadronic interaction Lagrangians}

The following Lagrangians describe the hadronic vertices.

\textit{$NN\eta'$ Lagrangian:}
\begin{equation}
{\cal L}_{NN\eta^\prime} =  - g_{NN\eta^\prime}
\bar N \left\{  \gamma_5\left[ i\lambda +
\frac{1 - \lambda}{2m_N}\, \fs{\partial}
\right]\eta^\prime\right\} N
~,
\label{NNeta'}
\end{equation}
where $\fs{\partial} = \gamma_\mu\partial^\mu$.

\textit{$NN^*\eta'$ Lagrangian (resonance $N^*=S_{11},P_{11}$): }
\begin{align}
{\cal L}^{(\pm)}_{NN^*\eta^\prime}
&=
\mp g_{NN^*\eta^\prime}
\bar N^*  \left\{ \Gamma^{(\pm)} \left[ i\lambda +
\frac{1 - \lambda}{m_N^*\pm m_N}\, \fs{\partial}
\right]\eta^\prime\right\} N
\nonumber\\
&\qquad +  \hc
\label{NN^*eta'}
\end{align}
where the upper and lower signs on the left refer to the even ($+$) and odd ($-$) parity
resonances, respectively. Following Refs.~\cite{Nak1,Nak2}, each of the $NB\eta^\prime$
vertices obtained from Eqs.~(\ref{NNeta'}) and (\ref{NN^*eta'}) ($B=N,N^*$) is multiplied
by a phenomenological cutoff function
\begin{equation}
  G_B(x) =\frac{\Lambda_B^4}{\Lambda_B^4+\left( x-m_B^2 \right)^2} \ ,
  \label{eq:ffhadron}
\end{equation}
which is normalized to unity, i.e., $G_B(m^2_B)=1$. The variable $x$ is the squared
four-momentum of the intermediate off-shell baryon $B$, whose mass $m_B$ is equal to
either the nucleon mass $m_N$ or the mass of the resonance, $m_{N^*}$. The cutoff
$\Lambda_B=1200$\,MeV is taken as the same for all baryons. The parameters $\lambda
\equiv \lambda_{NB\eta'}$ in Eqs.~(\ref{NNeta'}) and (\ref{NN^*eta'}) describing the
mixing of pseudoscalar and pseudovector contributions and the coupling constants
$g_{NB\eta'}$ are individual fit parameters for each of the three baryon states
considered here. (As the subsequent discussion shows, the fits prefer couplings that are
almost entirely pseudovector for the nucleon, i.e., $\lambda_{NN\eta'}\approx 0$, and
almost entirely pseudoscalar for the resonances, i.e.,  $\lambda_{NN^*\eta'}\approx 1$.)

\textit{$vv\eta'$ Lagrangian (vector meson $v=\rho,\omega$):}
\begin{subequations}
\begin{align}
{\cal L}_{\rho\rho\eta^\prime} &=  - \frac{g_{\rho\rho\eta^\prime}}{2m_\rho}
\varepsilon_{\alpha\beta\nu\mu} (\partial^\alpha \vec{\rho}^{\,\beta}) \cdot
(\partial^\nu \vec{\rho}^{\,\mu}) \eta^\prime ~,
\\
{\cal L}_{\omega\omega\eta^\prime} &=  - \frac{g_{\omega\omega\eta^\prime}}{2m_\omega}
\varepsilon_{\alpha\beta\nu\mu} (\partial^\alpha \omega^\beta)
(\partial^\nu \omega^\mu) \eta^\prime ~,
\end{align}
\end{subequations}
where $\vec\rho^{\,\mu}$ and $\omega^\beta$ stand for the $\rho$ and $\omega$ meson fields,
respectively. Each of the resulting $vv\eta^\prime$ vertices is multiplied by a product
of form factors,
$\tilde{F}_v(q_1^2)\, F_v(q_2^2)$, where $q_1=p_1-p_1'$ and $q_2=p_2-p_2'$ (cf.\ last diagram
in Fig.~\ref{fig:ppetaprod}). The form factor
\begin{equation}
  \tilde{F}_v(q^2) =\left( \frac{\Lambda_v^2}{\Lambda_v^2-q^2} \right)^2
\end{equation}
associated with one of the intermediate off-shell vector mesons is the same as in
Eq.~(\ref{eq:dipoleFF}), with the same cutoff masses $\Lambda_v$, except for the
normalization point, consistent with the kinematics at which the coupling constants
$g_{vv\eta^\prime}$ are extracted. The $g_{vv\eta^\prime}$ are obtained from a systematic
analysis of the radiative decay of vector and pseudoscalar mesons based on $SU(3)$ symmetry
considerations in conjunction with vector-meson dominance arguments \cite{Nak1}. Hence,
there are no free \emph{independent} parameters for this vertex.

\textit{$NN^*\pi$ Lagrangian (resonance $N^*=S_{11},P_{11}$):}
\begin{equation}
{\cal L}^{(\pm)}_{NN^*\pi}  =  \mp
\frac{g_{NN^*\pi}}{m_{N^*}\pm m_N} \,
\bar N^* \Gamma^{(\pm)}  (\fs{\partial}
\vec{\pi})\cdot\vec\tau N
+\hc
~,
\label{NN^*pi}
\end{equation}
where $\vec{\pi}$ denotes the pion field. Again, each of the resulting $NN^*\pi$ vertices
is multiplied by a product of form factors, $G_{N^*}(x)G_\pi(q^2_\pi)$; $q_\pi$ is the
pion's four-momentum and $x$, as before, is the squared four-momentum of the
intermediate $N^*$ state. The form factor $G_{N^*}$ here is exactly the same as in
Eq.~(\ref{eq:ffhadron}) for $B=N^*$, with $\Lambda_{N^*}=1200$\,MeV. $G_\pi$ is the pion
form factor parametrization from the Bonn potential, with a cutoff-mass value of
900\,MeV. For this vertex, therefore, the coupling constant $g_{NN^*\pi}$ is the only
additional fit parameter.

\textit{$NN^*v$ Lagrangian (vector meson $v=\rho,\omega$; resonance $N^*=S_{11},P_{11}$):}
\begin{equation}
{\cal L}^{(\pm)}_{NN^*v}  =  \frac {g_{NN^*v}} {m_{N^*}+m_N}\,
\bar N^* \Gamma^{(\mp)} \sigma_{\mu\nu}
(\partial^\nu V^\mu) N +\hc
~,
\label{NN^*v}
\end{equation}
where $V^\mu=\vec{\rho}^{\,\mu}\cdot\vec{\tau}, \, \omega^\mu$. Each resulting vertex is
multiplied by a product of form factors, $G_{N^*}(x)F_v(q^2_v)$; $q_v$ is the vector
meson's four-momentum and $x$ has the same meaning as before. The parameters of
$G_{N^*}$ and $F_v$ are fixed already; the coupling constant $g_{NN^*v}$, therefore, is
the only fit parameter here.

All of the remaining \textit{$MNN$ vertices} (meson $M=\pi,\eta,\rho,\omega,\sigma,a_0$)
are parametrized as in the Bonn potential \cite{Bonn}. The only exceptions are the
values of the coupling constant $g_{NN\omega}=10$, the pseudoscalar-pseudovector (ps-pv)
mixing parameter $\lambda_{NN\pi}=0$, and the cutoff-mass value of 900\,MeV at the $NN\pi$
vertex used in the resonance and meson exchange currents (see discussion in
Refs.~\cite{Nak2,Nak4}).

Throughout this work, the widths of the nucleon resonances are fixed to be $\Gamma_{N^*}
= 150$\,MeV ($N^*=S_{11},\, P_{11}$). We neglect their energy dependence in order to keep
the analysis simple. Certainly, such a feature should be taken into account when aiming
at a more quantitative extraction of the resonance parameters with data more accurate
than what are available at present.

In the present work we restrict ourselves to contributions from $\pi$, $\rho$, and
$\omega$ meson exchanges in the resonance currents in describing the $pp\to
pp\eta^\prime$ reaction. Also, in contrast to Ref.~\cite{Nak1}, we omit the
$\sigma\eta\eta^\prime$-exchange current in the present work because it is much less
under control than the dominant $vv\eta^\prime$-exchange contribution and its inclusion
would introduce additional uncertainties in the model.

\subsection{Gauge-invariance preserving contact term}

Employing form factors for the $s$- and $u$-channel contributions to the photoproduction
amplitude containing an intermediate nucleon (see the first two diagrams in
Fig.~\ref{fig:etaphotoprod}) and allowing for pseudovector couplings in the $NN\eta'$
vertex in general destroys the gauge invariance of the production amplitude. Within the
present context of a model approach, to restore gauge invariance requires the
introduction of phenomenological contact-type currents.

Following here the prescription given in Refs.~\cite{hh97g,hhtree98}, there are two basic
contributions necessary to ensure gauge invariance for the present application to
$\gamma p\to p\eta'$. The first
contribution,
\begin{equation}
  j^\mu_{\textsc{kr}}=-e\,g_{NN\eta'}(1-\lambda_{NN\eta'})
                \frac{\gamma_5\gamma^\mu}{2m_N}\left[ G_N(s)-G_N(u) \right]~,
  \label{eq:jKR}
\end{equation}
corresponding to the Kroll-Ruderman current of pion photoproduction, cancels the
gauge-invariance-violating terms arising from using pseudovector couplings. The form
factors $G_N$ here correspond to Eq.~(\ref{eq:ffhadron}). The second gauge-invariance
preserving (GIP) contribution,
\begin{align}
  j^\mu_{\textsc{gip}}
  &=-e\,g_{NN\eta'}\,\gamma_5\,\frac{(2p+k)^\mu)}{s-m_N^2} \left[ G_N(s)-\hat{F} \right]
  \nonumber\\
  &\quad\mbox{}
  -e\,g_{NN\eta'}\,\gamma_5\,\frac{(2p'-k)^\mu)}{u-m_N^2} \left[ G_N(u)-\hat{F} \right]~,
  \label{eq:jGIP}
\end{align}
is necessary because our model employs form factors at the vertices. As far as
gauge invariance is concerned, the function $\hat{F}$ here is arbitrary. Analyticity, on
the other hand, requires that this current be free of singularities, i.e., it must be a
\emph{contact} current.
One of the simplest choices for $\hat{F}$ in the present context then is\footnote{%
   In Refs.~\cite{hh97g,hhtree98}, it was argued that, for simplicity, one should choose
   to describe $\hat{F}$ in terms of the existing form factors
   of the problem at hand. The most general ansatz then would be
   \[
   \hat{F}(s,u,t) = 1- \sum_{i,j,k} \alpha_{ijk} [G_s(s)]^i [G_u(u)]^j [G_t(t)]^k \ ,
   \]
   where $G_s$, $G_u$, and $G_t$ are the $s$-, $u$-, and $t$-channel form factors. The simplest
   choice that is free of singularities is then given by restricting the sum to $i,j,k=0,1$
   and putting $\alpha_{ijk}=(-1)^{i+j+k}$. Equation (\ref{eq:Fhat}) follows with $G_t\equiv 0$
   and $G_s=G_u=G_N$.}
\begin{equation}
  \hat{F}= 1-\left[ G_N(s)-1 \right]\left[ G_N(u)-1 \right]~.
  \label{eq:Fhat}
\end{equation}
This corresponds to the choice advocated in Ref.~\cite{dw2} on the grounds of crossing
symmetry. The resulting GIP current then is
\begin{align}
  j^\mu_{\textsc{gip}}
  &=-e\,g_{NN\eta'}\,\gamma_5\,  (2p+k)^\mu G_N(u) \frac{G_N(s)-1}{s-m_N^2}
  \nonumber\\
  &\quad\mbox{}
  -e\,g_{NN\eta'}\, \gamma_5\, (2p'-k)^\mu G_N(s)\frac{G_N(u)-1}{u-m_N^2} ~,
  \label{eq:jGIP1}
\end{align}
which evidently is free of any singularities. Adding the sum of the Kroll-Ruderman term
(\ref{eq:jKR}) and the GIP current (\ref{eq:jGIP1}) restores gauge invariance for the
present model; in Fig.~\ref{fig:etaphotoprod}, they correspond to the rightmost diagram
in the top row of diagrams.

\begin{table*}[t]
\begin{center}
\caption{Model parameters fitted to the $\gamma p \to  \eta' p$ and $p p \to p p
\eta^\prime$ data. The dipole form factor is used at the electromagnetic vertex in the
mesonic current [cf. Eq.~(\ref{eq:dipoleFF})]. Below, ``Bonn" indicates that the same
values as in the Bonn $NN$ potential B (Table A.1)~\protect\cite{Bonn} are used. ($\dagger$)
indicates that the values of $g_{NN\omega}=10$ and
$(\lambda_{NN\pi},\Lambda_{NN\pi})=(0,900\,\mbox{MeV})$ overwrite those of the Bonn
potential. The widths of the resonances $N^*=S_{11}, P_{11}$ are fixed to be
$\Gamma_{N^*}=150$\,MeV. Also, the pseudovector coupling ($\lambda=0$) is used at the
$NN^*\pi$ vertices. Values in boldface are not fitted. Column (a) includes only the
meson-exchange current (mec) and the $S_{11}$ resonance current. Adding either the
nucleonic (nuc) or a $P_{11}$ resonance current contribution produces the results of
columns (b) and (c), respectively. In (d), successively stronger (as indicated by the
values of the $g_{NN\eta'}$ coupling constant in square brackets) nucleonic contributions
are added to the $\text{mec}+S_{11}+P_{11}$ contribution.}
\begin{tabular}{l@{\qquad}r@{\qquad}r@{\qquad}r@{\qquad}r}
\hline\hline
Coupling constant               &   (a)   &    (b)    &   (c)  &  (d) \\
\hline\hline
Nucleonic current: & & & & \\
$(g_{NN\gamma},~\kappa_p)$      &          & ($\boldsymbol{e}$,~\textbf{1.793}) &          & ($\boldsymbol{e}$,~\textbf{1.793}) \\
$(g_{NN\eta^\prime},~\lambda)$  &          & (2.22,~0.05)     &          & ([\textbf{1}, \textbf{2}, \textbf{3}], \textbf{0}) \\
$\Lambda_N$~(MeV)              &          & \textbf{1200}      &          & \textbf{1200} \\
$MNN~ [M=\pi,\eta,\rho,\omega,\sigma,a_0]$ &     &Bonn &     &Bonn \\
\hline
Mesonic current: & & & & \\
$g_{\eta^\prime\rho\gamma}$     & \textbf{1.25}  & \textbf{1.25}  & \textbf{1.25}  & \textbf{1.25}  \\
$g_{\eta^\prime\omega\gamma}$   & \textbf{0.44}  & \textbf{0.44}  & \textbf{0.44}  & \textbf{0.44}  \\
$\Lambda_v$ (MeV)               &  1383  &  1253  &  1400  & [1286,~1257,~1225]  \\
$g_{\eta^\prime\rho\rho}$       & \textbf{4.94}  & \textbf{4.94}  & \textbf{4.94}  & \textbf{4.94}  \\
$g_{\eta^\prime\omega\omega}$   & \textbf{4.90}  & \textbf{4.90}  & \textbf{4.90}  & \textbf{4.90}  \\
$MNN ~[M=\rho,\omega ]^{(\dagger)}$ &Bonn &Bonn &Bonn &Bonn \\
\hline
$N^*=S_{11}$ current: & & & & \\
$m_{N^*}$ (MeV)                               & 1760        & 1536        & 1646        & [\textbf{1650},~\textbf{1650},~\textbf{1650}] \\
$(g_{NN^*\gamma}\,g_{NN^*\eta^\prime},~\lambda)$ & (0.68,~1.00) & (4.16,~1.00) & (3.56,~0.76) & [(2.22,~0.98),~(2.45,~1.00),~(2.61,~1.00)] \\
$\Lambda_{N^*}$ (MeV)                             & \textbf{1200}  & \textbf{1200}  & \textbf{1200}  & \textbf{1200}            \\
$g_{NN^*\pi}\,g_{NN^*\eta^\prime}$              &  3.62       & 16.34       & 11.11       & [2.62,~4.37,~4.77]      \\
$g_{NN^*\rho}\,g_{NN^*\eta^\prime}$             & $-0.49$     & $-2.25$     & 11.25       & [11.01,~7.23,~6.69]     \\
$g_{NN^*\omega}\,g_{NN^*\eta^\prime}$           &  0.24       &  7.75       & $-1.93$       & [$-14.44$,~$-5.16$,~$-2.04$]  \\
$MNN ~[M=\pi,\rho,\omega]^{(\dagger)}$ &Bonn &Bonn &Bonn &Bonn \\
\hline
$N^*=P_{11}$ current: & & & & \\
$m_{N^*}$ (MeV)                               &             &             & 1873        & [1870,~1849,~1852]       \\
$(g_{NN^*\gamma}\,g_{NN^*\eta^\prime},~\lambda)$ &             &             & (4.60,~0.82) & [(3.28,~0.97),~(1.88,~0.90),~(0.27,~0.97)] \\
$\Lambda_N$~(MeV)                             &             &             & \textbf{1200}  & \textbf{1200}            \\
$g_{NN^*\pi}\,g_{NN^*\eta^\prime}$              &             &             &  6.04       & [4.61,~6.98,~9.45]       \\
$g_{NN^*\rho}\,g_{NN^*\eta^\prime}$             &             &             & $-2.20$       & [$-6.05$,~$-4.99$,~$-4.71$]    \\
$g_{NN^*\omega}\,g_{NN^*\eta^\prime}$           &             &             & $-20.53$      & [$-28.69$,~$-32.24$,~$-28.35$] \\
$MNN~[M=\pi,\rho,\omega]^{(\dagger)}$ &     &     &Bonn &Bonn \\
\hline\hline
\end{tabular}
\label{tab1}
\end{center}
\end{table*}

\section{Results and Discussion}

The basic strategy of our model approach is to first fix the free parameters of the
photoproduction reaction and then go to the hadronic process to fix the remaining
parameters.

The results for coupling constants and resonance masses, etc., given here were obtained
by standard best-fit procedures. At present, however, the quality of the data is not good
enough to provide really stringent constraints for the fits. As discussed also in detail
below, in many instances, therefore, the parameters obtained here may be changed within
certain limits without affecting the overall quality of the fits. In this situation,
$\chi^2$ values for the fits carry little information and were omitted from the tables.

\begin{figure*}[t]
\includegraphics[width=\columnwidth,angle=0,clip]{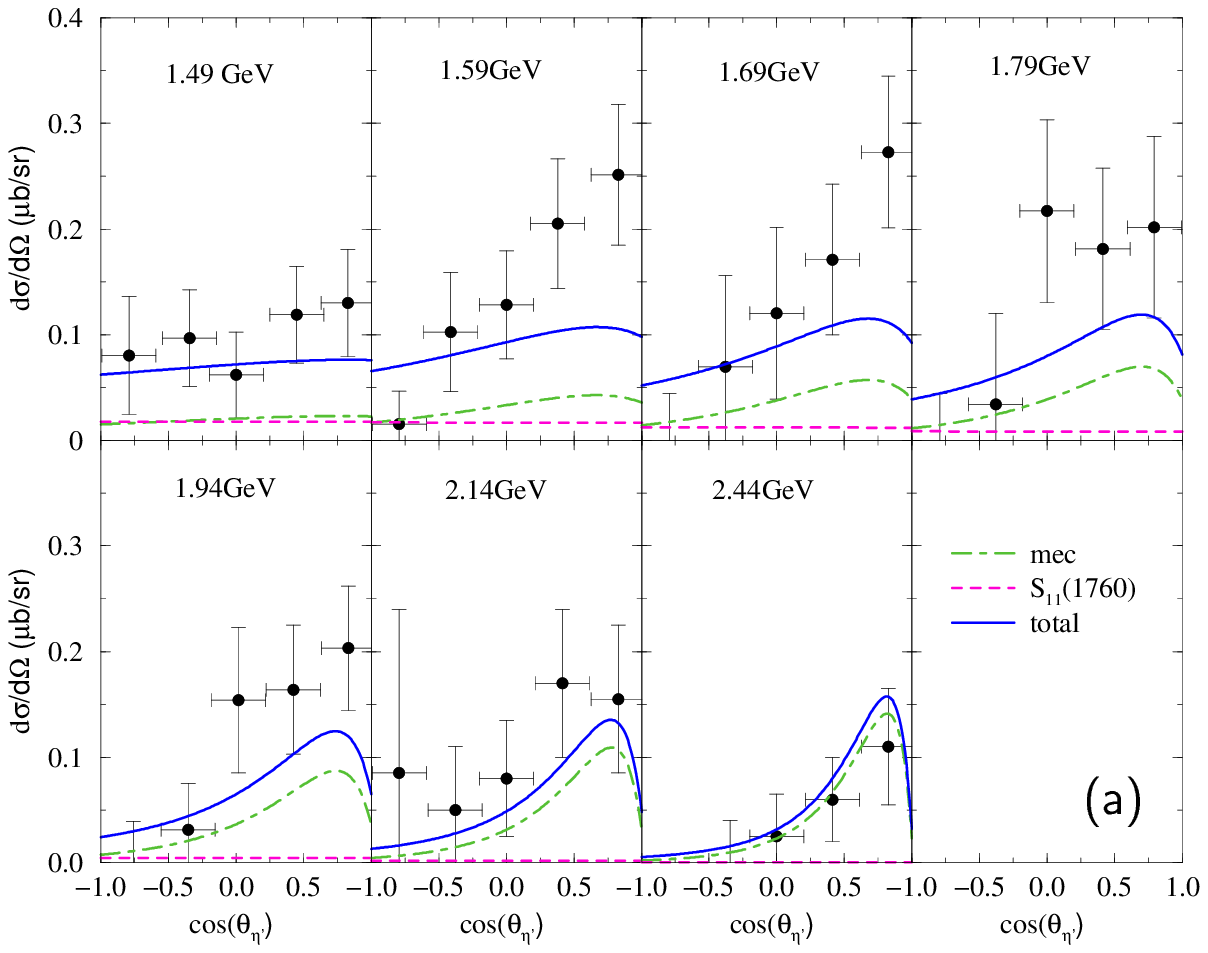}
\hfill
\includegraphics[width=\columnwidth,angle=0,clip]{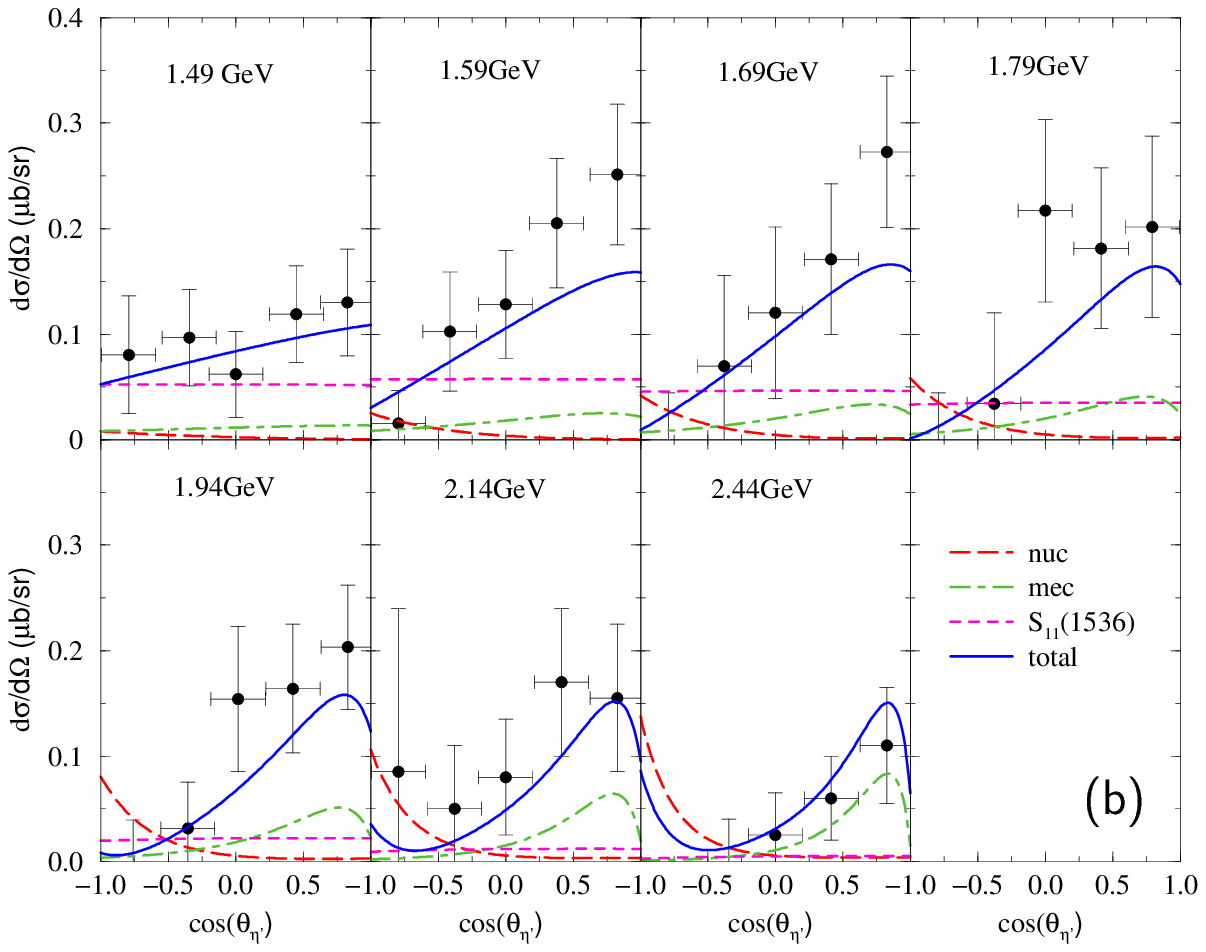}\\[2ex]
\includegraphics[width=\columnwidth,angle=0,clip]{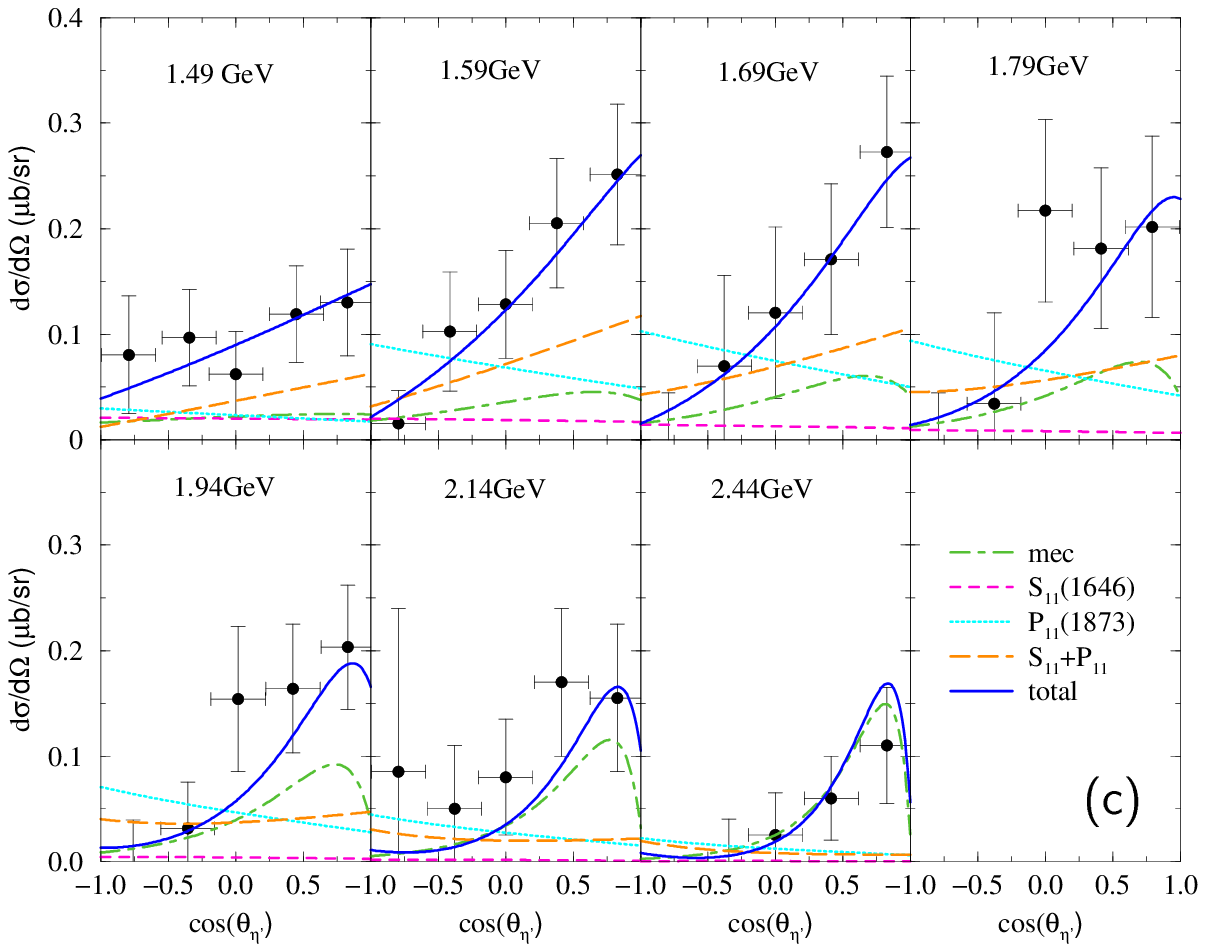}
\hfill
\includegraphics[width=\columnwidth,angle=0,clip]{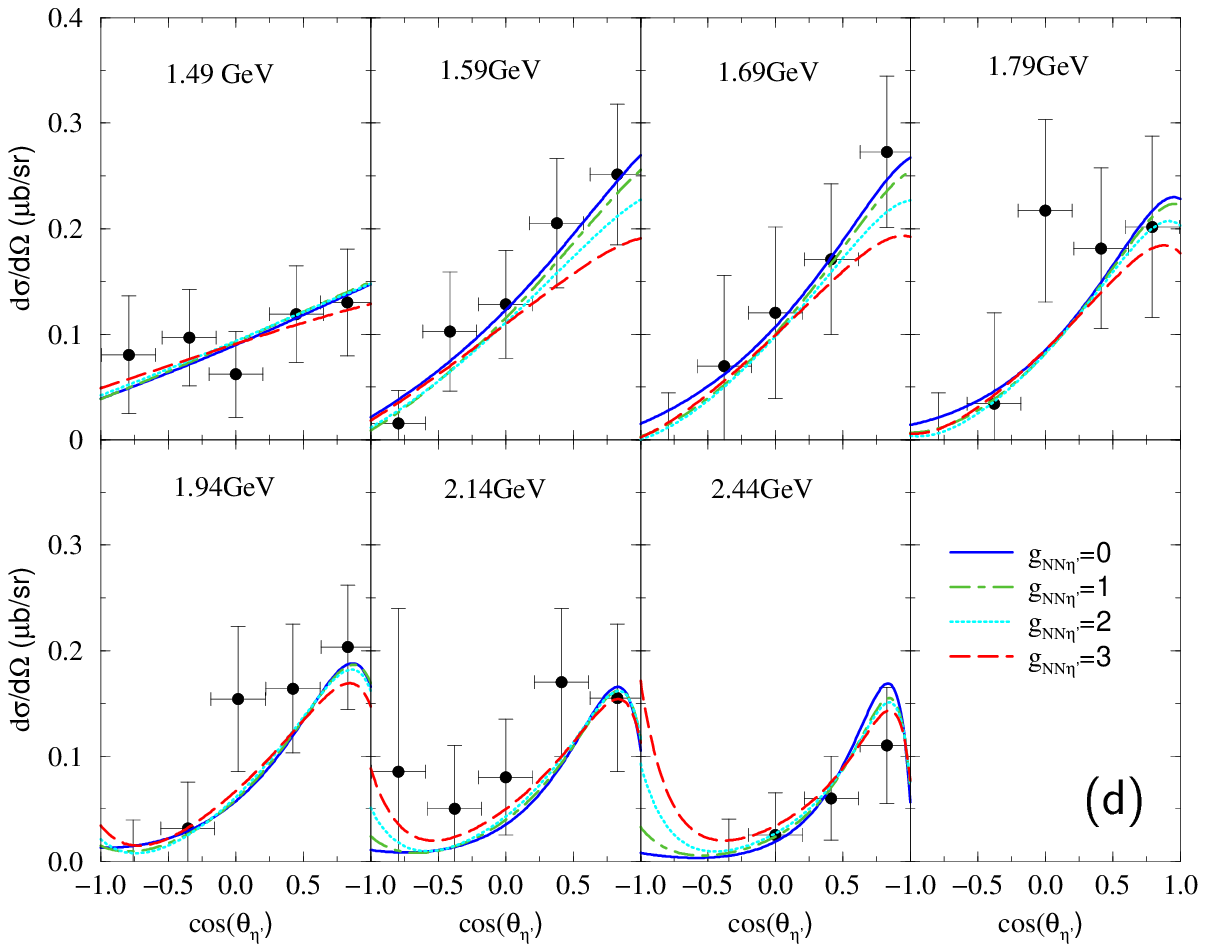}
\caption{\label{fig:photodiff}%
(Color online)
Differential cross section for $\gamma p\to p \eta'$ according to the mechanisms shown in
Fig.~\ref{fig:etaphotoprod}. Panel (a) includes only the meson-exchange current (mec) and
the $S_{11}$ resonance. Adding either the nucleonic (nuc)  contribution or a $P_{11}$
resonance produces the results of panels (b) and (c), respectively. In (d), successively
stronger (as indicated by the values of the $g_{NN\eta'}$ coupling constant) nucleonic
contributions are added to the results shown in panel (c). In each case, the model
parameters are determined by best fits. The meaning of the corresponding lines is
indicated in the panels. The data are from Ref.~\protect\cite{SAPHIR}.}
\end{figure*}

\begin{figure}[t]
\includegraphics[width=\columnwidth,angle=0,clip]{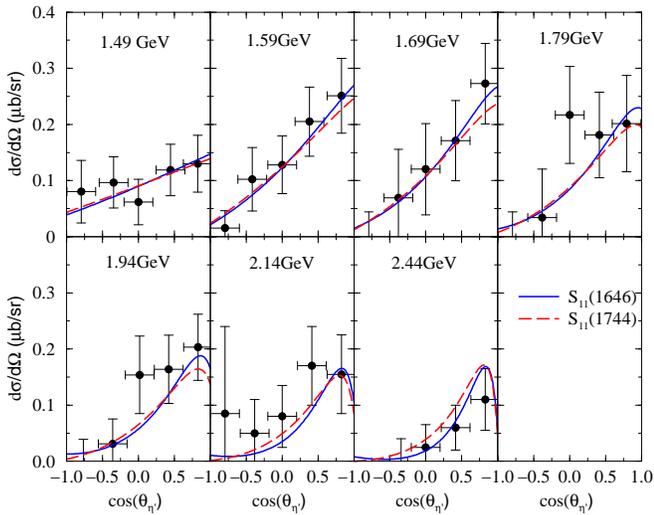}
\caption{\label{fig:resmass}%
(Color online)
Two fits resulting in different sets of the extracted resonance mass values. Both fits
include the $S_{11}$ and $P_{11}$ resonances as well as the meson-exchange currents. The
solid curves are the same ones shown in Fig.~\ref{fig:photodiff}(c) with the mass values
of $(m_{S_{11}}, m_{P_{11}})=(1646, 1873)$\,MeV. For the dashed curves, the corresponding
mass values are $(1744, 1879)$\,MeV.}
\end{figure}

\begin{turnpage}
\begin{figure*}
\includegraphics[height=\textheight,angle=270,clip]{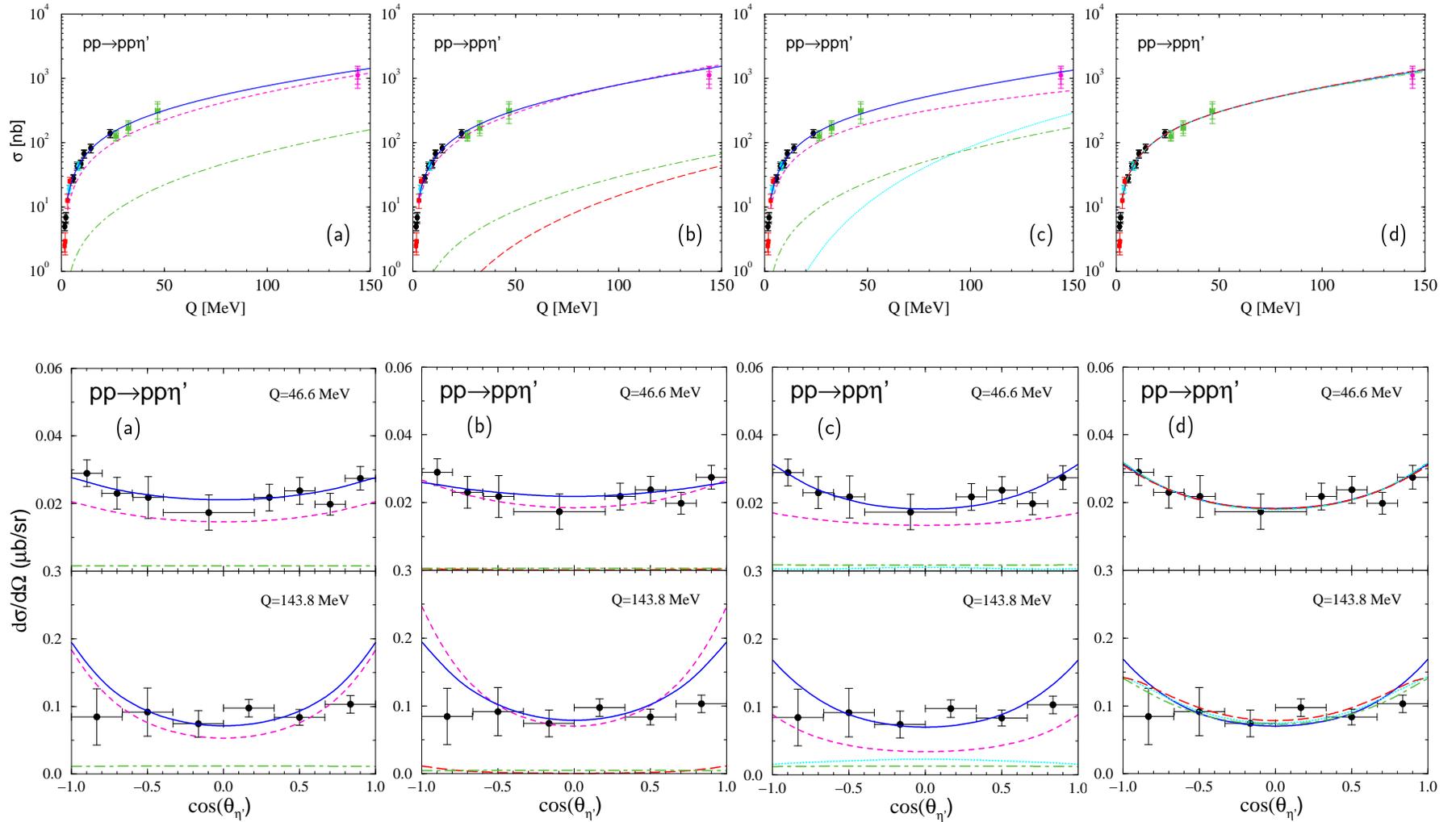}\\[4ex]
\caption{\label{fig:totang2}%
(Color online)
Excess energy, $Q$, dependence of the total cross section (top row of diagrams) and
angular distributions at $Q=46.6$ and 143.8 MeV in the c.m. frame of the system
(bottom row) for $pp\to pp \eta'$, according to the mechanisms depicted in
Fig.~\ref{fig:ppetaprod}. The panels labeled (a)--(d) in both rows correspond to the
respective panels (a)--(d) in Fig.~\ref{fig:photodiff}, and all line styles are explained
there. In part (d) of the total cross section and in the corresponding 47-MeV angular
distribution, on the present scales, all curves practically lie on top of each other,
i.e., these results are very insensitive to the nucleonic contributions. The total cross
sections comprise data from Refs.~\protect\cite{ppetapdata,Balestra,Khoukaz}; the
angular distribution data are from the COSY-11 collaboration (47 MeV) \protect\cite{Khoukaz}
and from DISTO  (144 MeV) \protect\cite{Balestra}. The calculations shown here incorporate
\emph{all three\/} data sets in the determination of the hadronic resonance coupling parameters
(in contrast to the results shown in Fig.~\ref{fig:totang1} below). }
\end{figure*}
\end{turnpage}

\begin{figure*}
\parbox{.6\textwidth}{%
\includegraphics[width=.6\textwidth,angle=0,clip]{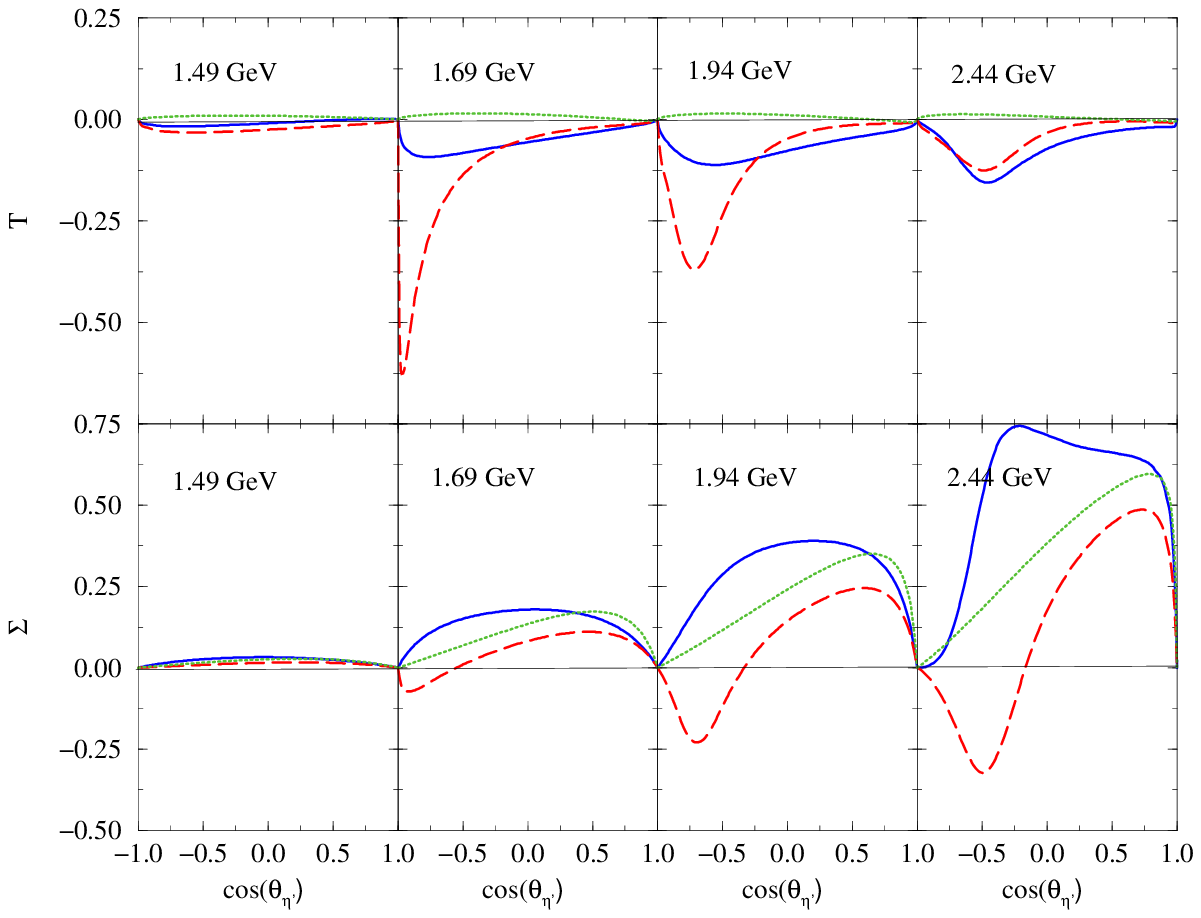}}
\hfill
\parbox{.37\textwidth}{%
\includegraphics[width=.35\textwidth,angle=0,clip]{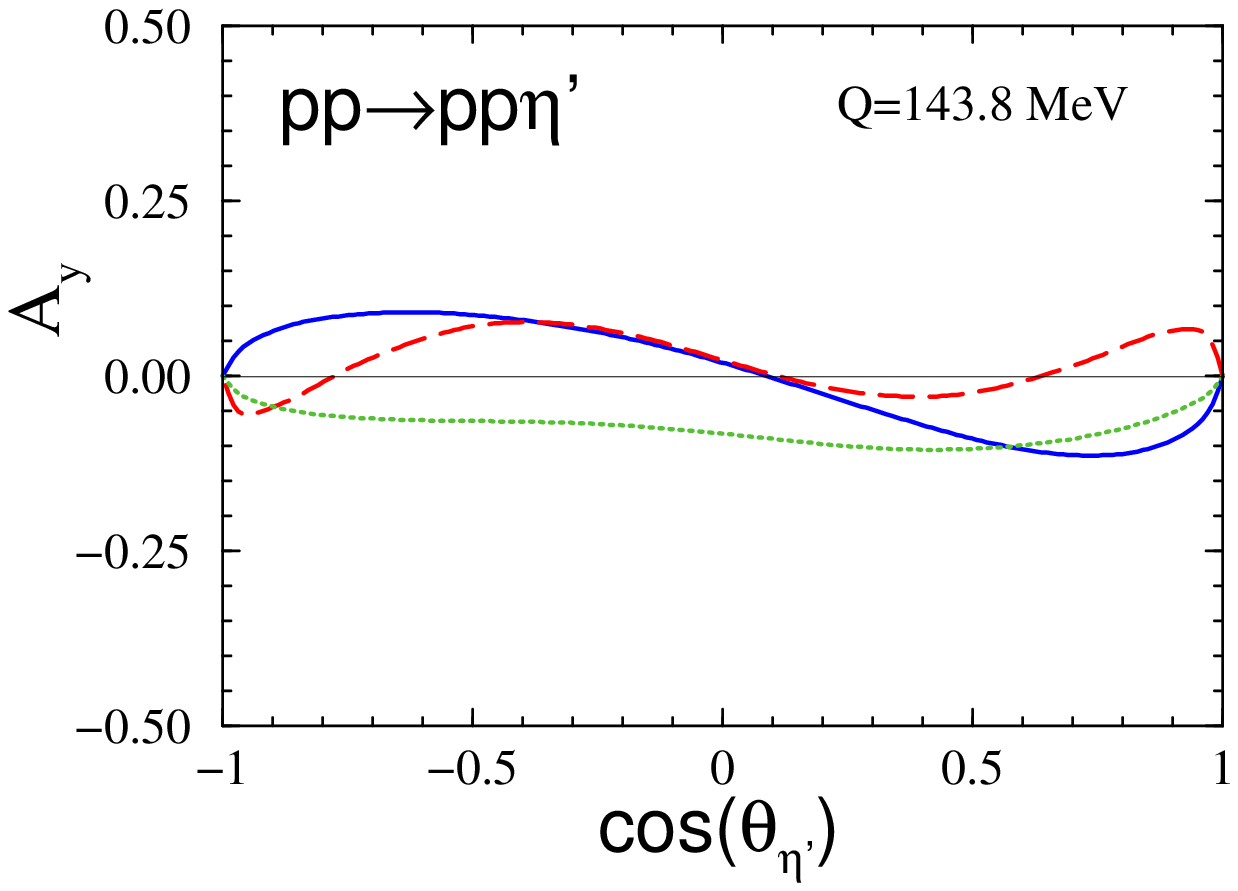}
\caption{\label{fig:spin2}%
(Color online)
Target and photon asymmetries $T$ and $\Sigma$, respectively, for $\gamma p\to p \eta'$
and analyzing power $A_y$ for $pp\to pp\eta'$. The solid lines correspond to the
parameters for the (d) panels in Figs.~\ref{fig:photodiff}--\ref{fig:totang2}, with
$g_{NN\eta'}=0$. The dashed lines are obtained when $g_{NN\eta'}=3$. The dotted
lines correspond to the parameters for the (a) panels in
Figs.~\ref{fig:photodiff}--\ref{fig:totang2}, where both the $P_{11}$ and nucleonic currents
are absent.}}
\end{figure*}

The steep rise and fall of the measured total cross section in $\gamma p \rightarrow p
\eta^\prime$ close to threshold \cite{ABBHHM,SAPHIR} suggests the presence of an $S_{11}$
nucleon resonance contribution. Of course, one should always keep in mind that there is
also the possibility of a threshold cusp effect, as discussed in Ref.~\cite{Hoehler},
that might explain the observed behavior of the photoproduction total cross section close
to threshold in the absence of any resonance. This requires further careful
considerations. Here we assume the observed behavior of the cross section to be due to
the nucleon resonance. Therefore, we first consider the $N^*=S_{11}$ resonance current.
The mass of the resonance as well as the product of the coupling constants,
$g_{NN^*\gamma}g_{NN^*\eta^\prime}$, and the ps-pv mixing parameter at the
$NN^*\eta^\prime$ vertex are free parameters to be fitted to the measured angular
distributions from $1.49$ to $2.44$\,GeV \cite{SAPHIR}. In addition to the $S_{11}$
resonance, we also consider the $\rho$ and $\omega$ meson-exchange currents in the
$t$-channel. For this current, the coupling constants at the production vertices
$\eta'v\gamma$ ($v=\rho, \omega$) are known from the radiative decay of $\eta^\prime$,
$\eta^\prime \rightarrow v + \gamma$. Also, since the relevant hadronic vertices $NNv$
are known from other studies, the only unknown parameter in the mesonic current is the
cutoff parameter $\Lambda_v$ in the form factor at the $\eta'v\gamma$ vertex [cf.\
Eq.~(\ref{eq:gammaVeta})]. Together with the free parameters of the $S_{11}$ resonance,
it has also been fitted to the data. The resulting parameter values are quoted in
Table~\ref{tab1}(a), and the corresponding angular distributions in
Fig.~\ref{fig:photodiff}(a). Here, the mass of the $S_{11}$ resonance results to be
$m_{S_{11}}=1760$\,MeV; however, inclusion of other currents into the fitting procedure
will change its value as we shall show below. As one can see from the figure, the
$S_{11}(1760)$ current contribution decreases as the energy increases while the mesonic
current contribution increases with the energy and rises at forward angles. At lower
energies, the constructive interference between the two currents is important in
enhancing the cross section, although it is not sufficient to reproduce the forward rise
exhibited by the data. At higher energies, the mesonic current dominates almost
completely and describes nicely the observed angular distribution. Therefore, the mesonic
current is fixed to a large extent by the forward angle data at higher energies.

Fig.~\ref{fig:photodiff}(b) shows the results when the nucleonic current is added to the
$S_{11}$ resonance and mesonic currents. In the nucleonic current, both the
$NN\eta^\prime$ coupling constant $g_{NN\eta^\prime}$ and the corresponding ps-pv mixing
parameter $\lambda_{NN\eta'}$ are fitted to the data. The parameters in the mesonic and
$S_{11}$ resonance currents are refitted to the data altogether. The corresponding values
are given in Table~\ref{tab1}(b). The nucleonic current contribution (long-dashed curves)
is small at lower energies but increases with energy at backward angles due to the
$u$-channel diagram, a feature that has been also realized in Ref.~\cite{Sibirtsev}.
Therefore, measurements at high energy and backward angles (large $t$) will help
constrain the poorly known $NN\eta^\prime$ coupling constant $g_{NN\eta^\prime}$. We will
come back to further discussion of this issue later. The $S_{11}$ resonance contribution
(dashed curves) is larger than that shown in Fig.~\ref{fig:photodiff}(a) at lower
energies which improves the description of the data in this energy region. It also
exhibits a stronger energy dependence. The fitted value of the resonance mass is now
$m_{S_{11}}=1536$\,MeV. A comparison with the value of $m_{S_{11}}=1760$\,MeV obtained in
Fig.~\ref{fig:photodiff}(a) illustrates how this parameter value changes with the
inclusion of different production mechanisms. Due to a constructive interference between
the nucleonic plus resonance current and the mesonic current in the forward angle region
at higher energies, the latter contribution is somewhat smaller than in
Fig.~\ref{fig:photodiff}(a). The overall description of the data is improved with respect
to that in Fig.~\ref{fig:photodiff}(a).

Fig.~\ref{fig:photodiff}(c) illustrates the effect of the $P_{11}$ resonance in $\gamma p
\rightarrow p \eta^\prime$. The mass of the $P_{11}$ resonance as well as the product of
the coupling constants, $g_{NN^*\gamma}g_{NN^*\eta^\prime}$, and the ps-pv mixing
parameter at the $NN^*\eta^\prime$ vertex ($N^*=P_{11}$) are free parameters to be fitted
to the data together with the parameters in the other currents. The resulting parameter
values are given in Table~\ref{tab1}(c). Here, the nucleonic current has been switched
off [it will be considered in the results shown in Fig.~\ref{fig:photodiff}(d)]. As can
be seen, the $P_{11}$ resonance contribution (dotted curves) rises at backward angles; it
also rises and falls with energy. The $S_{11}$ resonance contribution (dashed curves) is
relatively small, but its interference with the $P_{11}$ resonance contribution results
in a total resonance current contribution (long-dashed curves) that rises at forward
angles. Again, the mesonic current (dash-dotted curves) dominates at higher energies. The
solid curves correspond to the total contribution. The overall agreement with the data is
excellent, showing that the $P_{11}$ resonance may be required for a quantitative
description of the data. For a more definite conclusion about the role of the $P_{11}$
resonance more accurate data are called for. The fitted masses of the $S_{11}$ and
$P_{11}$ resonances are $m_{S_{11}}=1646$\,MeV and $m_{P_{11}}=1873$\,MeV, respectively.
Since an excellent agreement with the data is achieved at this point, we might identify
the $S_{11}$ resonance with the known $S_{11}(1650)$ resonance \cite{PDG}, whose quoted
width is $\Gamma_{S_{11}}=180$\,MeV. (Recall that, in this work, we have used a constant
width of $\Gamma_{N^*}=150$\,MeV for all the resonances.) The $P_{11}$ resonance does not
correspond to any known resonance; it is tempting to identify it with one of the missing
resonances with $m_{P_{11}}=1880$\,MeV and with the corresponding width of
$\Gamma_{P_{11}}=155$\,MeV, predicted by quark models \cite{Capstick1,Capstick2}.
Recently, an evidence for this resonance has been found in a three-channel unitary model
analysis \cite{Zagreb}. However, we emphasize that such an identification from the
present analysis is premature as we shall show later in connection with the results in
Fig.~\ref{fig:resmass}.

The role of the nucleonic current is illustrated in Fig.~\ref{fig:photodiff}(d). Here,
each curve corresponds to a given value of the $NN\eta^\prime$ coupling constant
$g_{NN\eta^\prime}$ as indicated. The pseudovector coupling ($\lambda_{NN\eta'}=0$) is
adopted. We mention that we have also considered the pseudoscalar coupling
($\lambda_{NN\eta'}=1$), but the fits do not support this choice and prefer to have a
small value of $\lambda_{NN\eta'}$ close to zero.\footnote{If we choose pure pseudoscalar
     coupling ($\lambda_{NN\eta'}=1$), we find that there is a strong transition between
     the positive and negative energy components of the nucleon wave functions in the final
     and intermediate states, respectively, since the intermediate nucleon is far off-shell
     owing to the fact that the present reaction involves the production of a massive
     particle ($\eta'$). As a result, the nucleonic current contribution becomes large for
     this choice of the coupling and this is not supported by the data.} For each value of
$g_{NN\eta^\prime}$, the parameters of the resonance and mesonic currents have been
refitted to best reproduce the data; the values are given in Table~\ref{tab1}(d). Here,
the mass of the $S_{11}$ resonance is fixed at $1650\ $\,MeV. The solid curves
corresponding to the choice $g_{NN\eta^\prime}=0$ are the same results as shown in
Fig.~\ref{fig:photodiff}(c). As one can see again, the major effect of the nucleonic
current shows up at higher energies and backward angles in the photoproduction reaction.
Accurate measurements in these kinematic regions are called for. In any case, judging
from the overall results, the existing data do not support values much larger than
$g_{NN\eta^\prime}=3$. In fact, they seem to prefer smaller value of $g_{NN\eta^\prime}$,
compatible with 0. This is considerably smaller than the value of
$g_{NN\eta^\prime}=6.1$ used in our previous work \cite{Nak1} on $pp \to pp\eta^\prime$,
and it is more in line with estimates based on the dispersion method \cite{Grein}.

As discussed in the Introduction, such a small value of the $NN\eta^\prime$ coupling
would have an important implication in connection to the ``spin puzzle'' of the nucleon.
A recent estimate \cite{Feldmann1} of $g_{NN\eta^\prime}$ based on an alternative formula
to Eq.~(\ref{spinfrac}) (neglecting the higher excited pseudoscalar states or glueballs
which are assumed to be negligible) in conjunction with the measured value of the singlet
axial charge yields a value of $g_{NN\eta^\prime}(0) = 1.4 \pm 1.1$. It should be noted
that the $NN\eta^\prime$ coupling constant entering in Eq.~(\ref{spinfrac}) is at zero
momentum squared, $g_{NN\eta^\prime}(q^2=0)$, while the coupling constant
$g_{NN\eta^\prime}$ in the present work is defined at the on-shell momentum squared,
$q^2=m_{\eta^\prime}^2$. We emphasize that the relatively small value of
$g_{NN\eta^\prime}$ found here is a model-dependent result. In particular, what is
relevant in our model is the product of the $NN\eta^\prime$ coupling constant and the
corresponding off-shell form factor. Since the intermediate nucleon in the nucleonic
current is far off-shell due to the large mass of the produced $\eta^\prime$ meson, the
result is sensitive to the choice of the form factor. As mentioned in the preceding
section, the form factor used at the $NN\eta^\prime$ vertex is the same as that used
consistently in our investigation of other meson production processes.

In Fig.~\ref{fig:resmass} we show a comparison of two fit results for $\gamma p
\rightarrow p \eta^\prime$, in which both the $S_{11}$ and $P_{11}$ resonance currents
were considered in addition to the meson exchange current. The solid curves are the same
results shown in Fig.~\ref{fig:photodiff}(c) with the resulting resonance masses of
$m_{S_{11}}=1646$\,MeV and $m_{P_{11}}=1873$\,MeV. The dashed curves correspond to a fit
with the resulting resonance masses of $m_{S_{11}}=1744$\,MeV and $m_{P_{11}}=1879$\,MeV.
For this fit, the values of the remaining fit parameters are very close to the
corresponding values quoted in Table~\ref{tab1}(c), except for the value of
$g_{NN^*\gamma}g_{NN^*\eta^\prime}=2.07$ for $N^*=S_{11}$ which compensates for the
change in the value of $m_{S_{11}}$. As one can see, the quality of the fit is
essentially the same in both cases; yet, the extracted values of $m_{S_{11}}$ differ by
$\sim 100$\,MeV from each other. (See also the results of a Regge trajectory calculation
in Fig.~\ref{fig:regge_resmass}, where the mass of the $P_{11}$ resonance varies
substantially.) This illustrates the order of uncertainties involved in the
identification of the resonances from the differential cross section data only (at least
from those currently available and from the type of analysis employed here). For a more
definitive identification of the resonances, one probably needs more exclusive data than
the cross sections, such as the spin observables shown in Fig.~\ref{fig:spin2}, which can
impose more stringent constraints. A further investigation of this issue is certainly
required.

The results for the $pp \rightarrow pp\eta^\prime$ reaction are shown in
Fig.~\ref{fig:totang2}. The two panels labeled (a) include the $S_{11}(1760)$ resonance
and the mesonic currents; they correspond to the photoproduction results of
Fig.~\ref{fig:photodiff}(a). All the relevant parameters for the latter reaction are
taken over unchanged. Thus, the additional parameters to be fitted for the hadronic
reaction concern the three products of the coupling constants,
$g_{NN^*\pi}g_{NN^*\eta^\prime}$, $g_{NN^*\rho}g_{NN^*\eta^\prime}$, and
$g_{NN^*\omega}g_{NN^*\eta^\prime}$ [where $N^*=S_{11}(1760)$], corresponding to the
three mesons exchanged between the two interacting nucleons in the resonance current (see
Fig.~\ref{fig:ppetaprod}). The resulting values are given in Table~\ref{tab1}(a). As can
be seen here, the dominant contribution is the $S_{11}(1760)$ resonance current (dashed
curves). The mesonic current is relatively small. The total cross section is nicely
reproduced, as well as the measured angular distribution at $Q=47$\,MeV.\footnote{The
          excess energy $Q$ is defined as $Q \equiv \sqrt{s}-\sqrt{s_o}$, where
          $\sqrt{s}$ denotes the total energy of the system and $\sqrt{s_o}=2m_N +
          m_{\eta^\prime}$ its $\eta^\prime$-production threshold energy.}
The latter exhibits some angular dependence although it might
be compatible with a flat shape within the given experimental uncertainties. The
completely flat angular distribution measured at $Q=144$\,MeV, however, is not
reproduced. As one can see, the calculated angular dependence is introduced by the
$S_{11}$ resonance and it arises due to the recoil of this resonance in the overall
center-of-mass (c.m.) frame.

Fig.~\ref{fig:totang2}(b) shows the results for $pp \rightarrow pp\eta^\prime$ which
include the nucleonic, $S_{11}(1536)$ resonance, and mesonic currents. Some of the
parameters are fixed from the photoproduction reaction corresponding to
Fig.~\ref{fig:photodiff}(b). As before, the remaining parameters are fitted to the $pp
\rightarrow pp\eta^\prime$ data and are given in Table~\ref{tab1}(b). Here, for the
purpose of consistency, one could, in principle, employ the coupling constants at the
$MNN^*$ vertex ($M=\pi, \eta, \rho, \omega$) for $N^*=S_{11}(1535)$ resonance as
determined from our recent study of the $pp \to pp\eta$ reaction \cite{Nak3}. We would
then have the coupling constant $g_{NN^*\eta^\prime}$ as the only free parameter to be
fitted. However, we have opted not to do so because Ref.~\cite{Nak3} did not aim for a
quantitative determination of those coupling constants. As can be seen here, the
$S_{11}(1536)$ resonance current (dashed curves) gives nearly the whole contribution to
the cross sections. The mesonic current is small followed by the nucleonic current. This
is in contrast to the results in our previous work \cite{Nak1}, where due to the scarcity
of the then available data, it had not been possible to constrain the individual current
contributions. The solid curves correspond to the total contribution. They exhibit
similar features to those shown in Fig.~\ref{fig:totang2}(a). Here, the model tends to
overestimate the total cross section at high energies although it still lies within the
experimental uncertainties. We may conclude, therefore, that the addition of the
nucleonic current at this stage does not improve the agreement with the data.

\begin{turnpage}
\begin{figure*}
\includegraphics[height=\textheight,angle=270,clip]{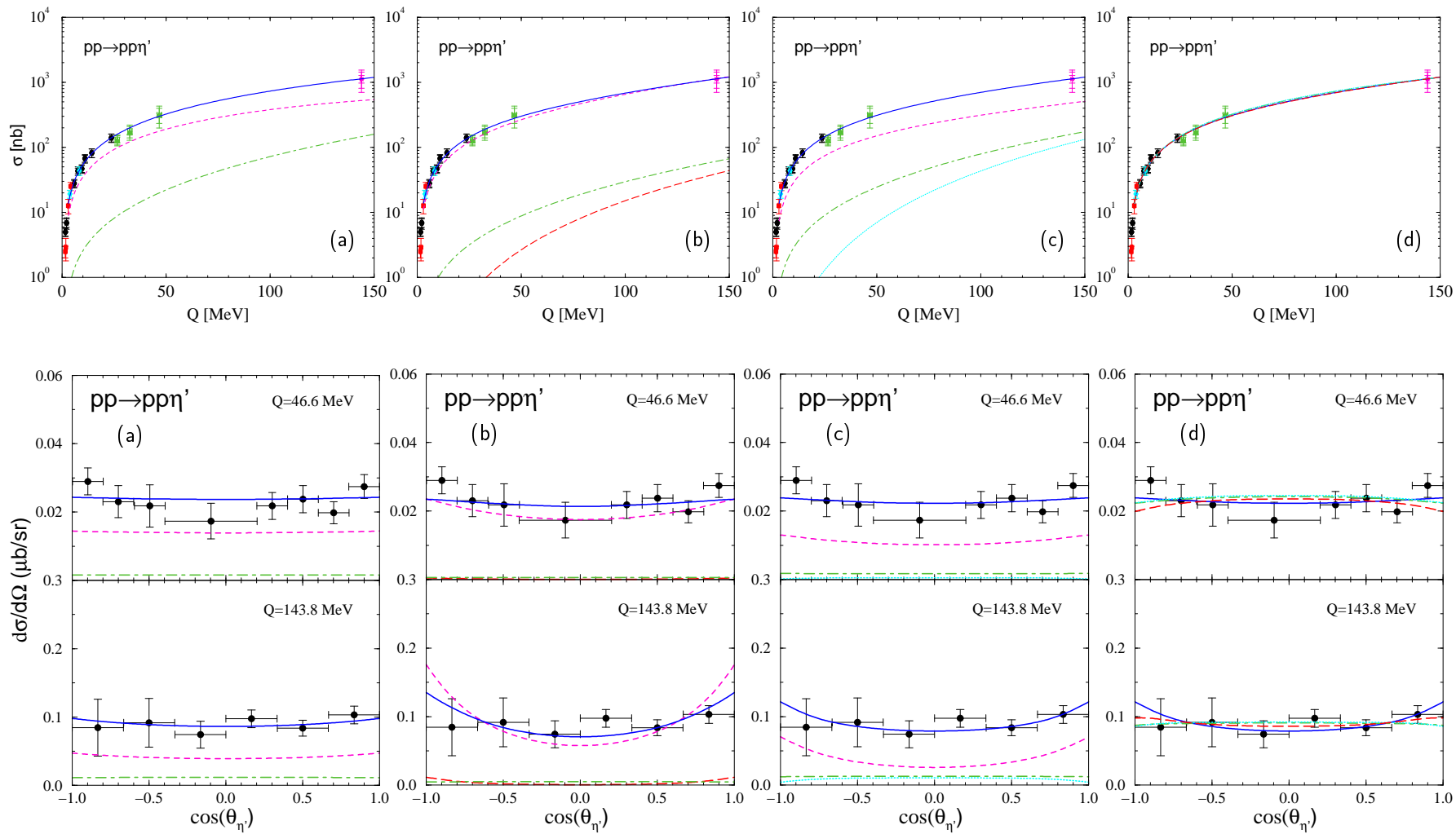}\\[4ex]

\caption{\label{fig:totang1}%
(Color online)
Same as Fig.~\ref{fig:totang2}, except that now the 47-MeV angular distribution data set
is excluded when fitting the hadronic parameters.}
\end{figure*}
\end{turnpage}

Next, we add the $P_{11}(1873)$ resonance to the mesonic and $S_{11}(1646)$ resonance
contributions. The results are shown in Fig.~\ref{fig:totang2}(c). Again, all parameters
relevant for the corresponding photoproduction reaction results of
Fig.~\ref{fig:photodiff}(c) are taken over. The additional parameters fitted for the
hadronic reaction are given in Table~\ref{tab1}(c). It is interesting to note that,
unlike in the photoproduction, here the $P_{11}(1873)$ contribution is much smaller than
that from $S_{11}(1646)$. The latter is the dominant current. The angular distribution at
$Q=144$\,MeV is somewhat improved; however, still in disagreement with the data. This
issue is further discussed in Fig.~\ref{fig:totang1}.

Fig.~\ref{fig:totang2}(d) illustrates the influence of the nucleonic current in the $pp
\rightarrow pp\eta^\prime$ reaction. The corresponding fitted parameters are given in
Table~\ref{tab1}(d). As one can see, this reaction is rather insensitive to the nucleonic
current contribution. This corroborates the statement in our earlier work \cite{Nak1}.

Fig.~\ref{fig:spin2} illustrates the sensitivity of some of the spin observables to the
coupling constant $g_{NN\eta^\prime}$ and also to the $P_{11}$ resonance. In the left
panel the target ($T$) and photon ($\Sigma$) asymmetries in the $\gamma p \rightarrow p
\eta^\prime$ reaction are shown. As can be seen, the target asymmetry is sensitive to
$g_{NN\eta^\prime}$ at backward angles around 1.69--1.94\,GeV (compare the solid and
dashed curves). However, this is the region where the cross section is very small. The
photon asymmetry becomes sensitive at higher energies and in a wider range of the
$\eta^\prime$ emission angle. This observable, therefore, may be helpful in constraining
$g_{NN\eta^\prime}$ more than just simple cross sections. The sensitivity to the $P_{11}$
resonance can be assessed by comparing the dotted (without $P_{11}$) and solid (with
$P_{11}$) curves. It is interesting to note that the influence of this resonance has a
different pattern than that of the nucleonic current. The right panel shows the
sensitivity of the analyzing power in the $pp \rightarrow pp\eta^\prime$ reaction.
Although the cross sections are rather insensitive to both $g_{NN\eta^\prime}$ and
$P_{11}$ resonance, this observable exhibits some degree of sensitivity.

\begin{table}[b!]
\begin{center}
\caption{Same as in Table~\ref{tab1}, except that here the angular distribution
data at $Q=46.6$\,MeV in $p p \to p p \eta^\prime$ were excluded from fitting.
Only those parameters affected by this exclusion in the fitting procedure are
displayed.}
\begin{tabular}{lrrrr}
\hline\hline
coupling constant               &   (a)   &    (b)    &   (c)  &  (d) \\
\hline\hline
$N^*=S_{11}$ current: & & & & \\
$g_{NN^*\pi}g_{NN^*\eta^\prime}$              & -0.03       & 13.33       & 10.02       & [0.05,0.98,3.28]      \\
$g_{NN^*\rho}g_{NN^*\eta^\prime}$             &  5.49       &  4.48       &  9.23       & [14.53,11.92,8.42]     \\
$g_{NN^*\omega}g_{NN^*\eta^\prime}$           &  1.27       &  2.30       &  7.41       & [2.25,6.31,4.25]  \\
\hline
$N^*=P_{11}$ current: & & & & \\
$g_{NN^*\pi}g_{NN^*\eta^\prime}$              &             &             &  3.92       & [13.01,1.54,16.47]       \\
$g_{NN^*\rho}g_{NN^*\eta^\prime}$             &             &             & -7.83       & [10.21,-3.90,8.81]    \\
$g_{NN^*\omega}g_{NN^*\eta^\prime}$           &             &             &  20.70      & [5.17,12.23,18.16] \\
\hline\hline
\end{tabular}
\label{tab2}
\end{center}
\end{table}

In order to investigate the discrepancy between our model results and the measured flat
angular distribution at $Q=144$\,MeV for $pp \to pp\eta^\prime$, we have repeated the
calculation shown in Fig.~\ref{fig:totang2} excluding the COSY-11 angular distribution
data at $Q=47$\,MeV from fitting. The resulting values of the hadronic couplings are
displayed in Table~\ref{tab2} and the corresponding cross sections in
Fig.~\ref{fig:totang1}. Although the total cross sections are reproduced with the same
quality as in Fig.~\ref{fig:totang2}, the angular distributions shown in panels (a)--(d)
at $Q=144$\,MeV are now much flatter and bring the model results in better agreement with
the data at this energy. Here, the less pronounced angular distribution is due to a
flatter $S_{11}$ resonance contribution which, in turn, is due to the change in the
excitation mechanism of this resonance, in particular, due to interference effects among
the exchanged meson ($\pi,\rho,\omega$) contributions in the $S_{11}$ resonance current.
This can be inferred from comparing the resulting coupling constants in Tables~\ref{tab1}
and \ref{tab2}. The predicted angular distributions at $Q=47$\,MeV in panels (a)--(d) are
now practically isotropic. They may be considered as being still compatible with the data
given the experimental error bars. However, the overall results in
Figs.~\ref{fig:totang2} and \ref{fig:totang1} may also indicate that the COSY-11 and
DISTO angular distribution data could be incompatible with each other. In this
connection, it is interesting to note that the angular distribution measured recently in
$pp\to pp\eta$ at $Q=41$\,MeV is completely isotropic \cite{Roderburg}. There, the
dominant $\eta$-production mechanism is the $S_{11}(1535)$ resonance current \cite{Nak3}.
Since in the present model, the dominant $\eta^\prime$-production mechanism is the
$S_{11}(1650)$ resonance current, it is natural to expect a similar feature for the
angular distribution in both reactions. Possible differences may, however, originate from
the eventual difference in the excitation mechanism of these two $S_{11}$ resonances as
mentioned above. For this purpose, it would be very interesting to measure other
observables, such as the invariant mass distribution, which are more sensitive to
excitation mechanisms of nucleon resonances \cite{Nak3}. In any case, independent
measurements of the angular distribution for energies $Q>50$\,MeV would help resolve the
issue as far as the shape of the angular distribution is concerned.

\begin{figure*}[t!]
\includegraphics[width=\columnwidth,angle=0,clip]{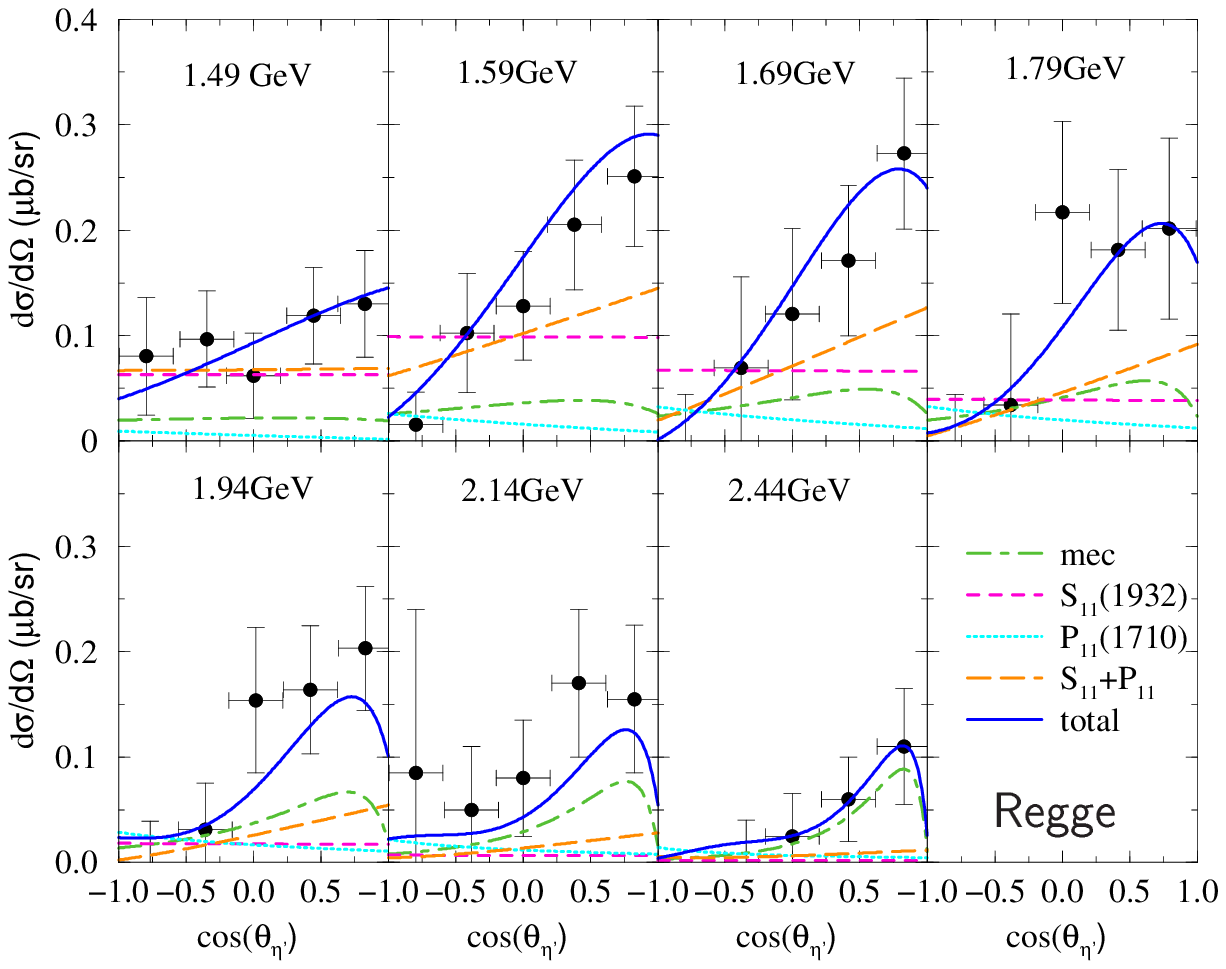}
\hfill
\includegraphics[width=\columnwidth,angle=0,clip]{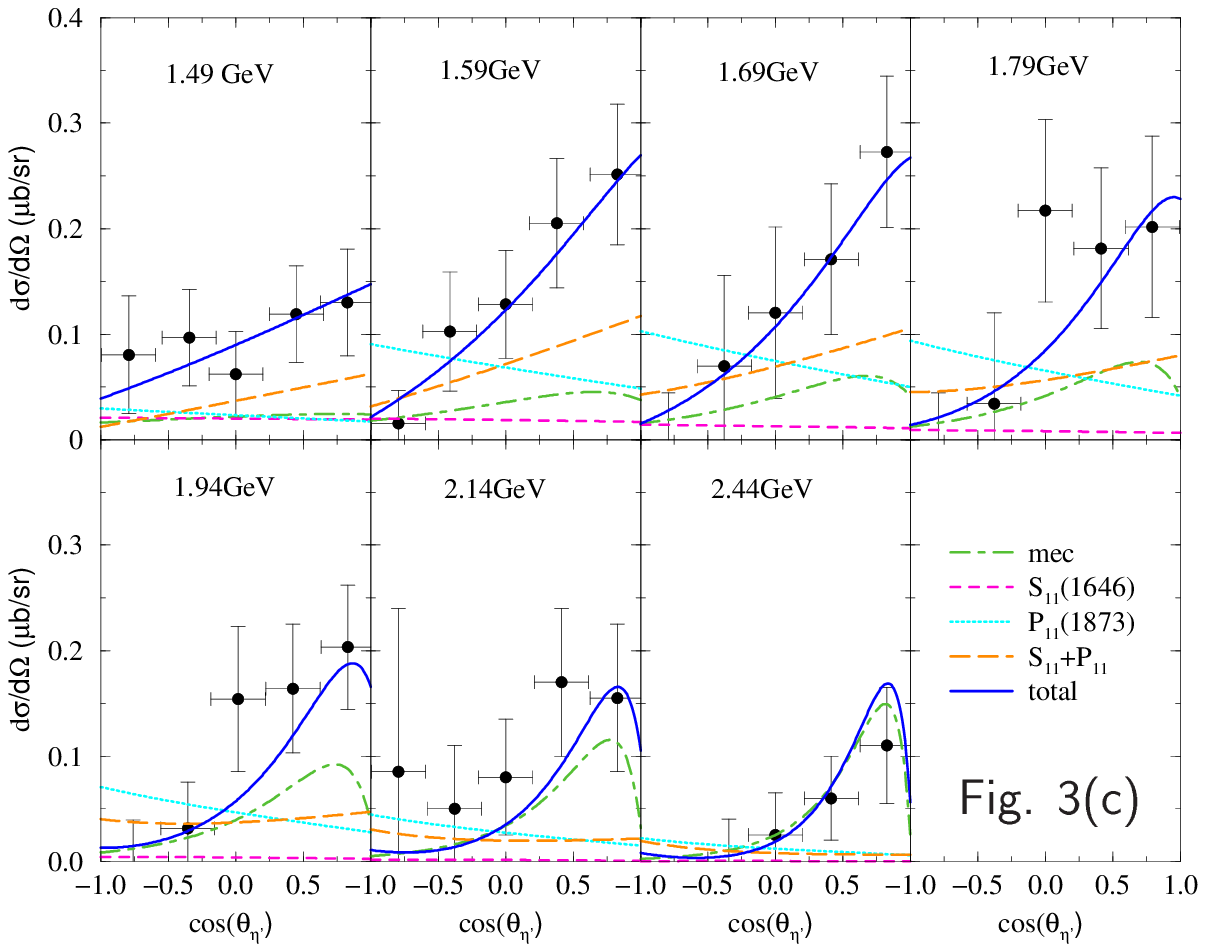}
\caption{\label{fig:regge}%
(Color online)
Comparison of differential cross sections for $\gamma p\to p \eta'$ using Regge
trajectories (left), similar to the treatment of Ref.~\protect\cite{Chiang}, and
conventional meson-exchange currents (right) for the $t$-channel vector meson exchange.
[The figure on the right-hand side is identical to Fig.~\ref{fig:photodiff}(c); it is
repeated here to allow for a better side-by-side comparison.]}
\end{figure*}

\subsection{Comparison to other approaches}

Recently, Chiang \textit{et al.} \cite{Chiang} have also investigated the $\gamma p
\rightarrow p \eta^\prime$ reaction. They concluded that the consideration of $t$-channel
vector-meson exchanges in terms of Regge trajectories is crucial in describing the
available data and that the $\rho$ and $\omega$ meson exchanges are unable to reproduce
the data. In contrast to the present work, they have not introduced a form factor at the
$\eta^\prime v\gamma$ vertex ($v=\rho , \omega$). In their calculations, the observed
forward rise of the cross section is due to the interference between the Regge and the
$S_{11}$ resonance contributions. Furthermore, no need for any $P_{11}$ resonance was
found in order to describe the data. In our opinion, the application of Regge
trajectories --- designed for high energies and low $t$ \cite{y1} --- in the low energy
regime of the SAPHIR data \cite{SAPHIR} is debatable. Moreover, such an approach cannot
be used (at least not straightforwardly) in the $pp \rightarrow pp\eta^\prime$ reaction
due to the important $NN$ FSI which has to be included in any model describing this
process. This involves a loop integration for which the Regge propagator as given in
Ref.~\cite{Chiang} cannot be used. The Regge theory is a theory designed for amplitudes.
Of course, with respect to our approach, one may criticize the use of a form factor at
the $\eta^\prime v\gamma$ vertex in the mesonic current, whereas we do not use any form
factor at any other electromagnetic vertex. However, the use of such a form factor may be
defended based on the results for the radiative decay process $\eta^\prime \rightarrow
\rho + \gamma \rightarrow \pi^+\pi^- + \gamma$. One may speculate that the relatively
strong form factor needed at the $\eta^\prime v\gamma$ vertex simulates effects of the
FSI ignored in the present approach.

\begin{table}[b!]
\begin{center}
\caption{Same as Table~\ref{tab1}(c), but using the exponential form factor
at the electromagnetic vertices in the mesonic current (column ``exp.'') and
using Regge trajectories (column ``Regge'').}
\begin{tabular}{lrr}
\hline\hline
coupling constant               &   exp.   &    Regge   \\
\hline\hline
Mesonic current: & & \\
$g_{\eta^\prime\rho\gamma}$     & {\bf 1.25}  & {\bf 1.25}  \\
$g_{\eta^\prime\omega\gamma}$   & {\bf 0.44}  & {\bf 0.44}  \\
$\Lambda_M (MeV)$               &   930       &              \\
$g_{\eta^\prime\rho\rho}$       & {\bf 4.94}  &             \\
$g_{\eta^\prime\omega\omega}$   & {\bf 4.90}  &             \\
$MNN [M=\rho,\omega ]^{(\dagger)}$ &Bonn &     \\
\hline
$N^*=S_{11}$ current: & & \\
$m_{N^*} (MeV)$                               & 1649        & 1932          \\
$(g_{NN^*\gamma}g_{NN^*\eta^\prime},\lambda)$ & (2.11,0.90) & (0.62,0.92)  \\
$\Lambda_N (MeV)$                             & {\bf 1200}  & {\bf 1200}    \\
$g_{NN^*\pi}g_{NN^*\eta^\prime}$              &  0.95       &               \\
$g_{NN^*\rho}g_{NN^*\eta^\prime}$             &  19.20      &               \\
$g_{NN^*\omega}g_{NN^*\eta^\prime}$           &  -35.46     &               \\
$MNN [M=\pi,\rho,\omega ]^{(\dagger)}$ &Bonn &    \\
\hline
$N^*=P_{11}$ current: & & \\
$m_{N^*} (MeV)$                               & 1874        & 1710          \\
$(g_{NN^*\gamma}g_{NN^*\eta^\prime},\lambda)$ & (4.03,0.85) & (5.93,0.58)  \\
$\Lambda_N (MeV)$                             & {\bf 1200}  & {\bf 1200}    \\
$g_{NN^*\pi}g_{NN^*\eta^\prime}$              &  8.91       &               \\
$g_{NN^*\rho}g_{NN^*\eta^\prime}$             & -29.44      &               \\
$g_{NN^*\omega}g_{NN^*\eta^\prime}$           & -32.26      &               \\
$MNN [M=\pi,\rho,\omega ]^{(\dagger)}$ &Bonn &     \\
\hline\hline
\end{tabular}
\label{tab3}
\end{center}
\end{table}

\begin{figure}[b!]
\includegraphics[width=\columnwidth,angle=0,clip]{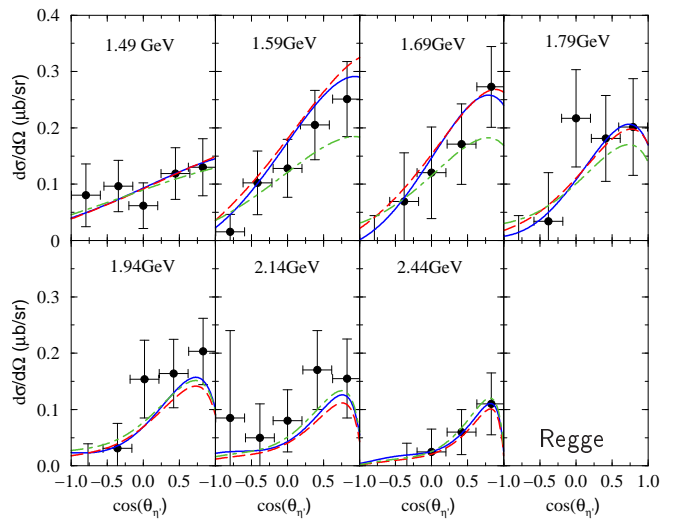}
\caption{\label{fig:regge_resmass}%
(Color online)
Three fits based on Regge trajectories resulting in different sets of the extracted
resonance mass values.
All the fits include the $S_{11}$ and $P_{11}$ resonances as well as the mec.
The solid curves are the same ones shown in Fig.~\ref{fig:regge}
with the mass values of $(m_{S_{11}}, m_{P_{11}})=(1932, 1710)$\,MeV.
For the dashed curves, the corresponding mass values are $(1932, 1950)$\,MeV.
The dash-dotted curves correspond to a fit with the mass values $(1650, 1811)$\,MeV.}
\end{figure}

In any case, we have also performed the calculation of the photoproduction reaction by
replacing the $t$-channel vector meson exchanges by the Regge trajectories following
Ref.~\cite{Chiang}. Apart from the obvious differences in the details of the treatment of
the resonance current, our calculation also differs in the sign of the
$\eta^\prime\omega\gamma$ coupling constant from that employed in Ref.~\cite{Chiang}. In
the present work the signs of the $\eta^\prime v\gamma$ couplings are inferred from a
systematic analysis \cite{Nak1} of the pseudoscalar and vector meson radiative decays
based on an $SU(3)$ Lagrangian in conjunction with the sign of the coupling constant
$g_{\pi v \gamma}$ determined from a study of pion photoproduction in the 1 GeV energy
region \cite{Garcilazo}. Our results using the Regge trajectories are shown in
Fig.~\ref{fig:regge}; the corresponding parameters are found in Table~\ref{tab3}. We see
that, overall, the results are basically the same as those of Fig.~\ref{fig:photodiff}(c)
using the conventional $\rho$ and $\omega$ meson exchanges. In particular, here also the
interference between the mesonic and resonance currents is the underlying mechanism
responsible for reproducing the observed angular distribution. This corroborates the
findings of Ref.~\cite{Chiang}. However, the resulting resonance masses of
$m_{S_{11}}=1932$\,MeV and $m_{P_{11}}=1710$\,MeV differ considerably from those obtained
in Fig.~\ref{fig:photodiff}(c). It is natural to ask whether this discrepancy is related
to the uncertainties in the determination of the resonance mass using only the cross
section data as illustrated in Fig.~\ref{fig:resmass} or whether it is due to different
approaches used in the treatment of the $t$-channel exchange contribution. To address
this question, in Fig.~\ref{fig:regge_resmass} we show the results of three different
fits using the Regge trajectories. The solid curves are the same results shown in the
left panel of Fig.~\ref{fig:regge}. The dashed curves correspond to another fit resulting
in the resonance masses of $(m_{S_{11}},m_{P_{11}})=(1932,1950)$\,MeV. One sees that the
qualities of both fits are comparable to each other but the corresponding values of
$m_{P_{11}}$ differ by more than $200$\,MeV, revealing once more (cf.\
Fig.~\ref{fig:resmass}) that the cross section data alone are insufficient to constrain
accurately the resonance masses. However, we were unable to fit the data (with a
comparable quality) with the mass of the $S_{11}$ resonance much smaller than
$m_{S_{11}}=1932$\,MeV. The dash-dotted curves in Fig.~\ref{fig:regge_resmass} correspond
to a fit with a fixed mass of $m_{S_{11}}=1650$\,MeV. This yields a fitted mass of
$m_{P_{11}}=1811$\,MeV for the $P_{11}$ resonance. This set of the resonance masses
$(m_{S_{11}},m_{P_{11}})=(1650,1811)$\,MeV is more in line with the set
$(1646,1873)$\,MeV obtained using the conventional vector meson exchanges in the
$t$-channel. Here, however, the quality of the fit is inferior to that achieved in
Fig.~\ref{fig:photodiff}(c), although the data are still reproduced within their
uncertainties. These considerations indicate that the determination of the resonance
masses is also quite sensitive to different approaches used. Further studies of this
issue are needed before a more unambiguous identification of the resonances can be made
from the $\eta^\prime$ photoproduction process. It should also be noted that our
calculations based on Regge trajectories differ in details from the results obtained in
Ref.~\cite{Chiang}. As mentioned above, these differences should, in part, be due to the
different treatment (in detail) of the resonance current contributions.

\begin{figure}
\centering
\includegraphics[width=\columnwidth,angle=0,clip]{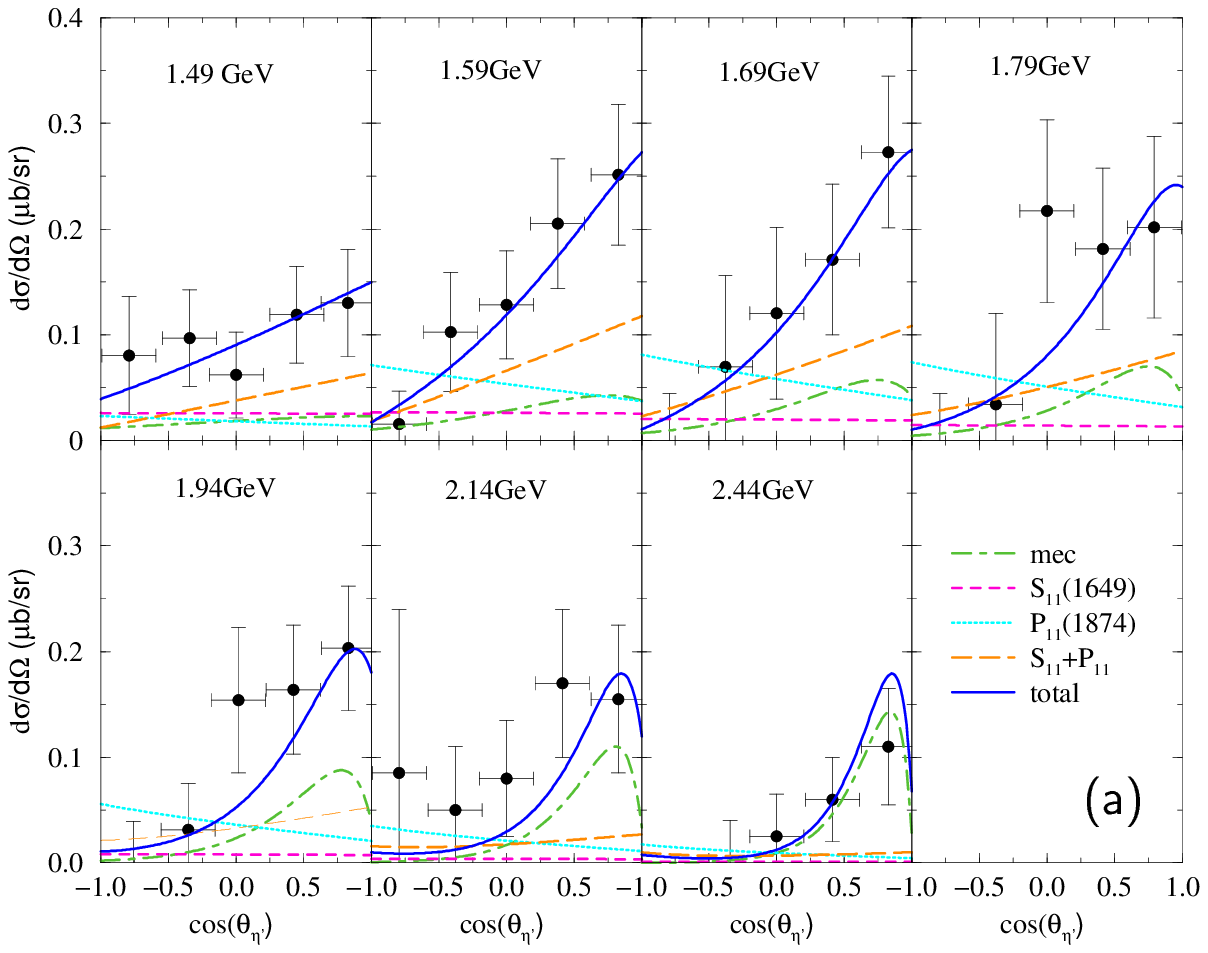}\\[3ex]
\includegraphics[width=.36\textwidth,angle=0,clip]{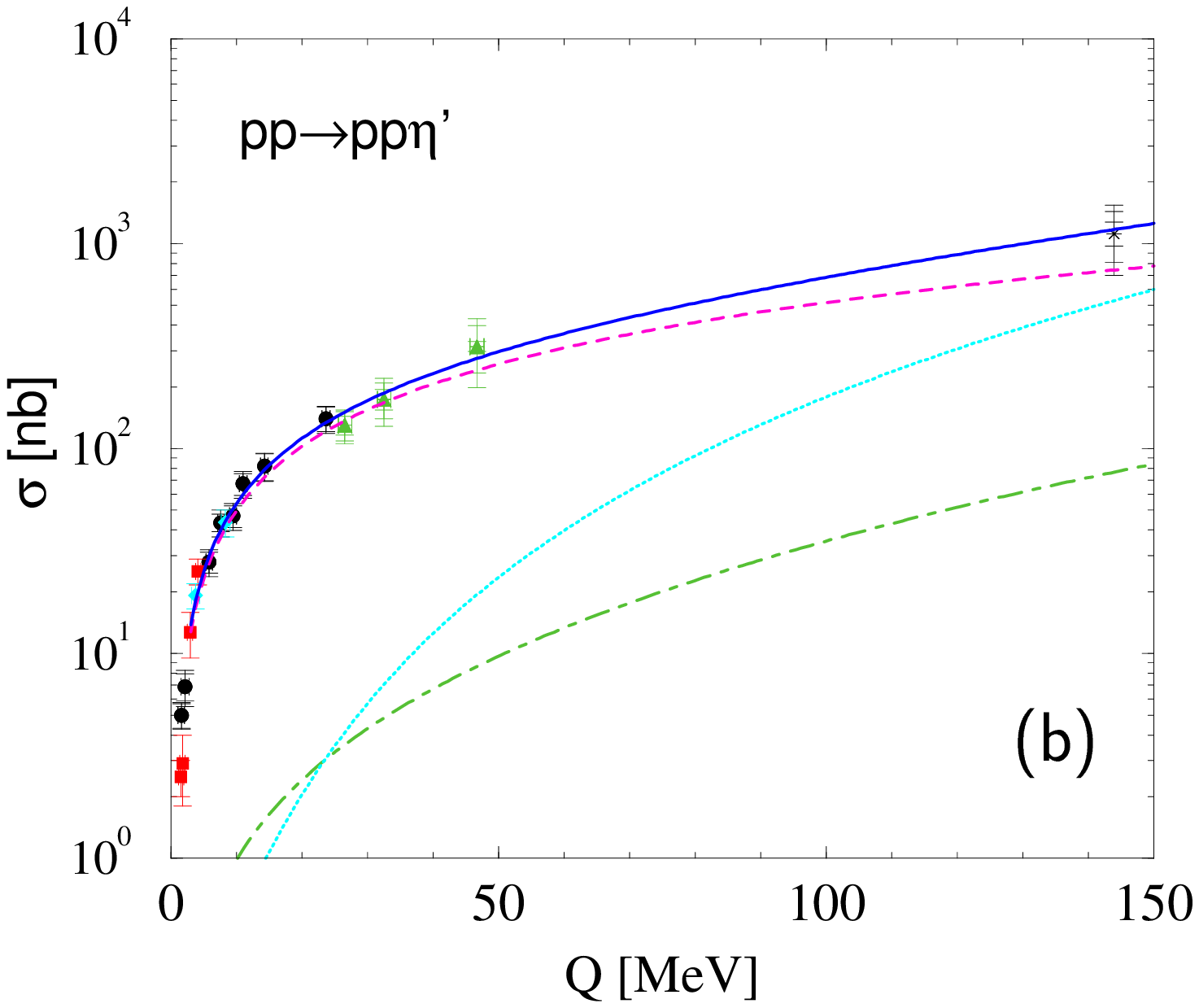}\\[2.5ex]
\includegraphics[width=.36\textwidth,angle=0,clip]{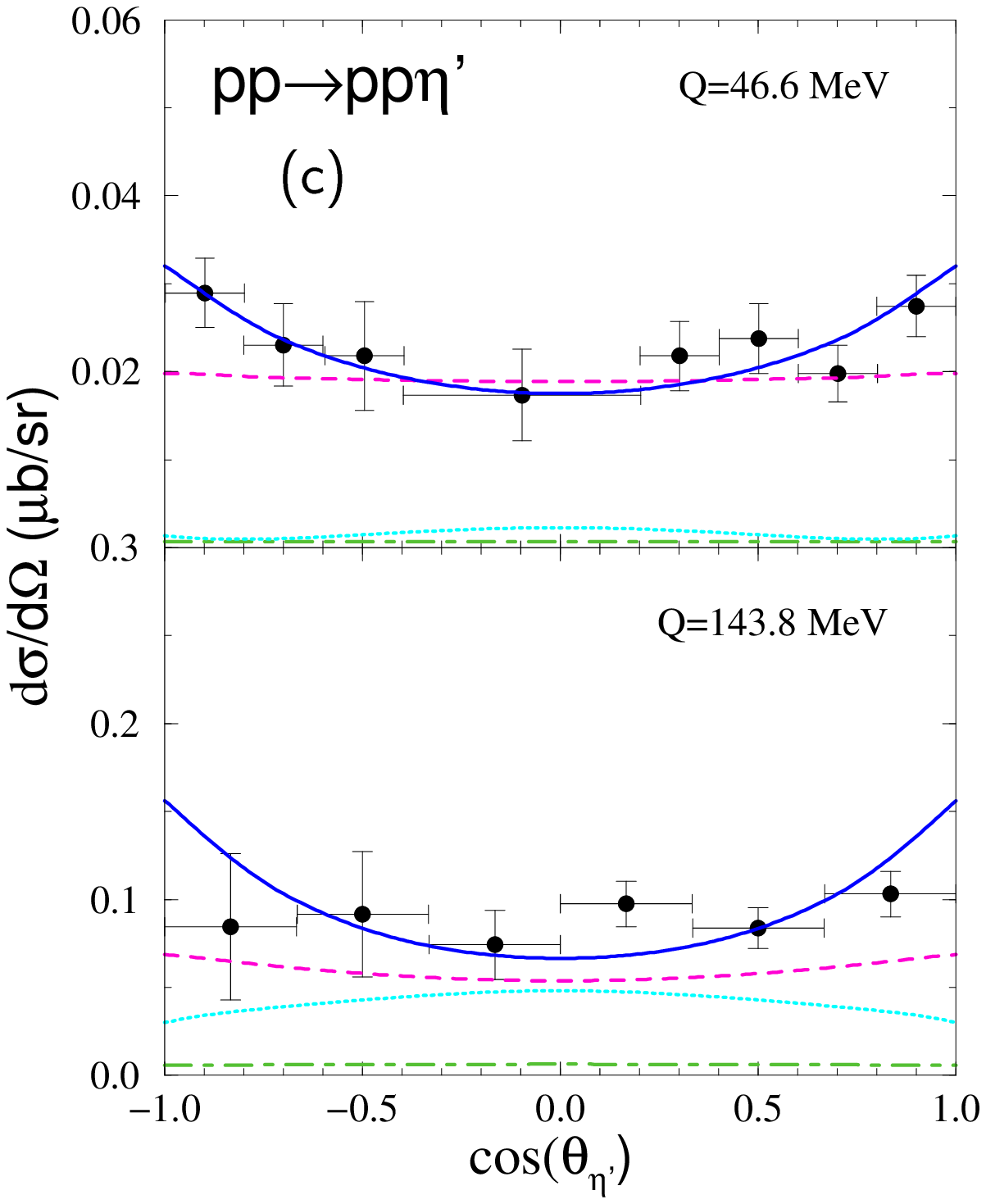}

\caption{\label{fig:e_mr2}%
(Color online) Fig.~(a) here corresponds to Fig.~\ref{fig:photodiff}(c), and
(b) and (c) correspond to the two respective panels labeled (c) in
Fig.~\ref{fig:totang2}, the difference being that now the $\gamma v \eta'$ vertex
functions of dipole form [see Eq.~(\ref{eq:dipoleFF})] have been replaced by exponential
form factors, $F_v(t)=\exp{\left[ (t-m_v^2)/\Lambda^2 \right]}$, similar to
Ref.~\protect\cite{Sibirtsev} (however, with a different normalization; see text). }
\end{figure}

Sibirtsev \textit{et al.} \cite{Sibirtsev} have also reported their study of the $\gamma
p \rightarrow p \eta^\prime$ reaction quite recently. In contrast to Ref.~\cite{Chiang}
and the present work, they describe the forward rise of the angular distribution
basically by the $\rho$ and $\omega$ meson exchanges in the $t$-channel. They achieve
this by including a ($t$-dependent) exponential form factor at the $\eta^\prime v\gamma$
vertex. Moreover, the $S_{11}(1535)$ resonance was introduced in order to describe the
steep rise and fall of the total cross section close to threshold. We have also repeated
our calculation employing an exponential form factor at the $\eta^\prime v\gamma$ vertex
instead of the dipole form factor; the corresponding parameters are given in
Table~\ref{tab3}. The results are shown in Fig.~\ref{fig:e_mr2}(a) which again exhibits
the same features observed in the calculation using the dipole form factor [see
Fig.~\ref{fig:photodiff}(c)]. We were not able to reproduce their results using only the
mesonic and $S_{11}(1535)$ resonance currents. The chief difference between the results
shown in Fig.~\ref{fig:e_mr2} and those in Ref.~\cite{Sibirtsev} is that we have used an
exponential form factor normalized to unity at the on-mass-shell point $q^2=m_v^2$,
consistent with the kinematics at which the coupling constant $g_{\eta^\prime v\gamma}$
is extracted. In Ref.~\cite{Sibirtsev}, the form factor is normalized at $q^2=0$ instead.
Figures~\ref{fig:e_mr2}(b) and \ref{fig:e_mr2}(c) show the corresponding results for $pp
\rightarrow pp\eta^\prime$ using the exponential form factor. They are essentially the
same as those using the dipole form factor.

The above considerations show that we arrive at the same conclusion, namely, that the
interference between the meson exchange and resonance currents is the mechanism
responsible for the angular distribution exhibited by the photoproduction data,
irrespective of whether one uses the $\rho$ and $\omega$ meson exchanges in the
$t$-channel (with either dipole or exponential form factor) or Regge trajectories.
Certainly, the problem of the Regge trajectory versus form factor is an extremely
important issue that needs to be addressed. Judging from our findings so far, it may well
be that both simulate the same physics not accounted for explicitly in these
calculations.

\section{Summary}

We have described consistently the $\gamma p \to p \eta^\prime$ and $pp\to pp\eta^\prime$
reactions within an approach based on a relativistic meson-exchange model of hadronic
interactions. The model includes the nucleonic and the mesonic, as well as the
nucleon-resonance currents. The photoproduction process is made gauge-invariant by adding
a phenomenological contact current that parametrizes the effect of the final-state
interactions. The $pp\to pp\eta^\prime$ is described within the distorted-wave Born
approximation in which both the initial and final state $NN$ interactions are taken into
account explicitly.

For $\eta^\prime$ photoproduction, we have shown that the mesonic as well as the $S_{11}$
and $P_{11}$ resonance currents are important to describe the existing data. Our
analysis, where the widths of the resonances were set to $\Gamma_{N^*}=150$\,MeV, yields
a position close to 1650 MeV  and 1870 MeV for the $S_{11}$ and $P_{11}$ resonances,
respectively. This suggests that the former resonance may well be identified with the
known $S_{11}(1650)$ resonance \cite{PDG}, whose quoted width is
$\Gamma_{S_{11}}=180$\,MeV. The $P_{11}$ resonance, in contrast, does not correspond to
any known resonance. It is tempting to identify it with one of the missing resonances
predicted at 1880 MeV with the correspond width of $\Gamma_{P_{11}}=155$\,MeV
\cite{Capstick1,Capstick2}. We emphasize, however, that one should be cautious with such
an identification of the resonances. As we have seen, the cross section data alone do not
impose enough constraints for an unambiguous determination of the resonances. To do so
probably requires more exclusive data than just the cross sections. Moreover, the
extracted values of the resonance masses are quite sensitive to the model used in the
description of the photoproduction process. In particular, the issue of Regge
trajectories versus conventional vector meson exchanges (with form factors) is of extreme
importance. These points require further investigation before a conclusive identification
of the resonances can be made.

Our study also shows that the nucleonic current should be relatively small. Indeed, the
available photoproduction data prefer this current to be compatible with zero. In any
case, the $NN\eta^\prime$ coupling constant cannot be much larger than
$g_{NN\eta^\prime}=3$. The $\eta^\prime$ photoproduction reaction may impose a more
stringent constraint on $g_{NN\eta^\prime}$, provided one measures the cross sections at
higher energies and backward angles. In this respect, as we have also shown, spin
observables such as the photon asymmetry might be suited better than the cross sections.
It should be noted that the result pertaining here to the $NN\eta^\prime$ coupling
constant is, of course, a model dependent one. Indeed, what is relevant in our
calculations is the product of $g_{NN\eta^\prime}$ and the associated form factor.

We have also addressed the contradictory conclusions as to the underlying reaction
mechanisms arrived in the recent work by two independent groups \cite{Chiang,Sibirtsev}.
In our consistent calculations, whether introducing a form factor at the electromagnetic
vertex in the $t$-channel meson-exchange current or using the Regge trajectories instead,
one arrives at the same conclusion; namely, the observed angular distribution is due to
the interference between the $t$-channel and the nucleon resonance $s$- and $u$-channel
contributions, irrespective of the particular approach one uses. It is conceivable,
therefore, that the phenomenological aspects of the various approaches (including the
present one) may be simulating the same physics not taken into account explicitly.

As for the $p p \to p p \eta^\prime$ reaction, the present study yields the $S_{11}$
resonance as the dominant contribution to the production current. The $P_{11}$ resonance,
mesonic and nucleonic currents are much smaller than the $S_{11}$ resonance current. The
combined analysis of this and the photoproduction reaction was crucial for these
findings. The details of the excitation mechanism of the $S_{11}$ resonance, however, are
not constrained by the currently existing data. To learn more about the relevant
excitation mechanism, observables other than the cross sections, such as the invariant
mass distribution, are necessary \cite{Nak3}, in addition to measuring the $p n \to p n
\eta^\prime$ and/or $p n \to d \eta^\prime$ process. This process will help disentangle
the isoscalar and isovector meson-exchange contributions.

The present model cannot describe the flat angular distribution in $p p \to p p
\eta^\prime$ measured by the DISTO collaboration \cite{Balestra} at $Q=144$\,MeV, once
the recently measured angular distribution by the COSY-11 collaboration \cite{Khoukaz} at
$Q=47$\,MeV is included in a global fitting for the relevant hadronic coupling
parameters. The calculated result exhibits a pronounced angular dependence. If we wish,
however, a flatter angular distribution at $Q=144$\,MeV --- compatible with the DISTO
data within the given experimental uncertainties --- can be achieved provided the
$Q=47$\,MeV angular distribution data is excluded from fitting. Doing so, the predicted
angular distribution at $Q=47$\,MeV comes out to be nearly completely isotropic. Although
this seems still compatible with the COSY-11 data, the latter exhibits some angular
dependence which is too disturbing to be ignored. Independent measurements of the angular
distribution for excess energies $Q > 50$\,MeV will help resolve this issue.

Finally, the results of the present work should provide useful information for further
investigations, both experimentally and theoretically, of the $\gamma N \to N
\eta^\prime$ and $NN \to NN \eta^\prime$ reactions.

\begin{acknowledgments}
The authors would like to thank A. Khoukaz and the COSY-11 collaboration for providing
the new COSY-11 data prior to publication. This work was supported by the COSY Grant No.\
41445282\,(COSY-58). The work of H.\,H. was supported in part by Grant No.\
DE-FG02-95ER-40907 of the U.S. Department of Energy.
\end{acknowledgments}

\end{document}